\documentclass[conference]{IEEEtran}

\usepackage{cite}
\usepackage{comment}

\ifCLASSINFOpdf
  \usepackage[pdftex]{graphicx}
  \graphicspath{{figures/}} 
\DeclareGraphicsExtensions{.pdf,.jpeg,.png,.eps}
\else
\fi
\usepackage{amsmath,amssymb}
\usepackage{xcolor}
\usepackage{listings}
\usepackage{multirow}
\usepackage{makecell}
\usepackage{indentfirst}
\usepackage{algorithm}
\usepackage{algpseudocode}
\usepackage{varwidth}
\algrenewcommand\algorithmicrequire{\textbf{Input:}}
\algrenewcommand\algorithmicensure{\textbf{Output:}}

\newcommand{\vk}{{\bf k}}
\newcommand{\vq}{{\bf q}}

\newcommand{\JY}[1]{{\color{red}{(JY: #1)}}}
\newcommand{\CY}[1]{{\color{blue}{(CY: #1)}}}

\begin{document}
\title{GPU Acceleration of Non-equilibrium Green's Function Calculation using OpenACC and CUDA FORTRAN}

% \author{\IEEEauthorblockN{}
% \IEEEauthorblockA{School of Electrical and\\Computer Engineering\\
% Georgia Institute of Technology\\
% Atlanta, Georgia 30332--0250\\
% Email: http://www.michaelshell.org/contact.html}
% \and
% \IEEEauthorblockN{Homer Simpson}
% \IEEEauthorblockA{Twentieth Century Fox\\
% Email: homer@thesimpsons.com}
% \and
% \IEEEauthorblockN{James Kirk\\ and Montgomery Scott}
% \IEEEauthorblockA{Starfleet Academy\\
% San Francisco, California 96678--2391\\
% Telephone: (800) 555--1212\\


\author{\IEEEauthorblockN{Jia Yin\IEEEauthorrefmark{1},
Khaled Z. Ibrahim\IEEEauthorrefmark{2},
Mauro Del Ben\IEEEauthorrefmark{2}, 
Jack Deslippe\IEEEauthorrefmark{3},
Yang-hao Chan\IEEEauthorrefmark{4} and
Chao Yang\IEEEauthorrefmark{2}}
\IEEEauthorblockA{\IEEEauthorrefmark{1}School of Mathematical Sciences,
Fudan University\\
Shanghai, China 200433\\ Email: jiayin@fudan.edu.cn (Jia Yin)}
\IEEEauthorblockA{\IEEEauthorrefmark{2}Applied Mathematics \& Computational Research Division (AMCRD),
Lawrence Berkeley National Laboratory\\
Berkeley, California 94720\\ Email: cyang@lbl.gov (Chao Yang), kzibrahim@lbl.gov (Khaled Z. Ibrahim), mdelben@lbl.gov (Mauro Del Ben)}
\IEEEauthorblockA{\IEEEauthorrefmark{3}National Energy Research Scientific Computing Center (NERSC), Lawrence Berkeley National Laboratory\\ Berkeley, California 94720\\ Email: jrdeslippe@lbl.gov (Jack Deslippe)}
\IEEEauthorblockA{\IEEEauthorrefmark{4}Institute of Atomic and Molecular Sciences, Academia Sinica\\ Taipei, Taiwan 10617\\ Email: yanghao@gate.sinica.edu.tw (Yang-hao Chan)}}


\maketitle

\begin{abstract}
The numerical solution of the Kadanoff-Baym nonlinear integro-differential equations, which yields the non-equilibrium Green’s functions (NEGFs) of quantum many-body systems, poses significant computational challenges due to its high computational complexity. In this work, we present efficient implementations of a numerical method for solving these equations on distributed-memory architectures, including many-core CPUs and multi-GPU systems.
For CPU-based platforms, we adopt a hybrid MPI/OpenMP programming model to exploit both inter-node and intra-node parallelism. On GPU-accelerated systems, we implement the method using two distinct approaches: MPI/OpenACC and MPI/CUDA FORTRAN. Several optimization strategies are employed to enhance GPU performance, including techniques to maximize computational resource utilization and minimize the overhead associated with kernel launches and memory management. Although OpenACC is easy to use, CUDA FORTRAN provides more advanced features for configuring and managing multiple levels of concurrency, while also simplifying memory allocation and data movement between host and device. This flexibility translates into significant performance improvements.
We compare the performance of the three implementations and demonstrate that the GPU-based approaches achieve substantial speedups over CPU-based implementations. Furthermore, both CPU and GPU versions exhibit excellent strong and weak scaling, confirming the scalability and efficiency of our approach for large-scale NEGF computations.
\end{abstract}


%\ifCLASSOPTIONpeerreview
%\begin{center} \bfseries EDICS Category: 3-BBND \end{center}
%\fi
\IEEEpeerreviewmaketitle

\section{Introduction}

\label{sec:intro}
%\begin{itemize}
%    \item Describe the problem to be solved
%    \item point out computational challenges
%    \item describe previously developed solver (programming model (MPI/OpenMP) and performance on CPUs
%    \item briefly describe the need to develop solver that can be run on GPUs and the new approach (OpenACC) taken in this work
%
%    \item Why is it difficult to use the concurrency and memory bandwidth on GPU?
%
%    \item The performance of OpenACC compared with CUDA Fortran
%\end{itemize}

Simulating the dynamics of a nonequilibrium quantum many-body system with a large number of particles $N$ is computationally challenging due to the exponential scaling of the number of degrees of freedom in the many-body wavefunction with respect to $N$. However, many properties of a quantum many-body system can be ascertained by working with its observables such as the single-particle Green's function $G(r,t,r',t')$, which can be interepreted as the probability of detecting a particle at the position $r$ and time $t$ when a particle is created at position $r'$ and time $t'$. For simplicity, we will leave out the spatial coordinates in $G$ below. For systems that are out of equilibrium, the non-equilibrium Green's function (NEGF) is a two-time function $G(t,t')$ that satisfies an equation of motion known as the Kadanoff-Baym equation~\cite{kadanoff1962quantum,stefanucci2013nonequilibrium,kadanoff2018quantum,FW}, which is a set of nonlinear integral differential equation defined on a contour $\mathcal{C}$ in the complex plane known as the Keldysh contour~\cite{BK,keldysh1965zhetf}. 

The Kadanoff-Baym equations (KBEs)~\cite{lipavsky1986generalized,kadanoff2018quantum} have the form
\begin{equation}\label{eq:kbe}
\left\{
    \begin{aligned}
    &\left[i\frac{d}{dt}-h(t)\right]G(t, t') =\delta(t,t') + \int_\mathcal{C}\Sigma(t, \overline{t})G(\overline{t}, t')d\overline{t},\\
    &\left[-i\frac{d}{dt'}-h(t')\right]G(t,t') = \delta(t,t') + \int_\mathcal{C}G(t, \overline{t})\Sigma(\overline{t}, t')d\overline{t},
    \end{aligned}
\right.
\end{equation}
where $h(t)$ is a single-particle Hamiltonian, and $\Sigma(t,t')$ is the self-energy that accounts for many-body interaction among different particles.  The true analytical form of $\Sigma(t,t')$ is generally unknown, and is often approximated through many-body perturbation theory.

The numerical solution of~\eqref{eq:kbe} requires propagating $G(t,t')$ on a two-time grid. The time evolution procedure is computationally expensive due to the coupled nature of the equations, the nonlinearity of integro-differential equation, and the need to evaluate the integral term at each time step in addition to the self-energy evaluations. These self-energy terms are dependent on the Green's function, and their evaluation requires several tensor contractions. 

Fortunately, multiple levels of concurrency exist in both the self-energy and collision integral calculations. The algorithmic concurrency allows us to implement these calculations efficiently on many-core CPUs using the MPI+OpenMP programming model. However, performing a long-time NEGF dynamics simulation on a many-core CPU based distributed memory high performance computer can still take a prohibitively long wall clock time.

In recent years, there has been widespread availability of high-performance computers equipped with graphics processing unit (GPU) accelerators. GPUs possess significantly more hardware threads than CPUs and offer much higher memory bandwidth. They appear to be the natural computing platform for carrying out NGEF simulations.

However, there are some unique features of the GPUs that make it challenging to take full advantage of the computational resources available on such an architecture.
To optimize the efficiency of numerical KBE solvers on multiple GPUs, several modifications of the MPI-OpenMP implementation of the KBE solver are required to maximize resource utilization and minimize computational overhead.

We discuss the changes we need to make in an MPI+OpenACC implementation of the self-energy and collision integral calculations in order to achieve satisfactory performance on NVIDIA GPUs. 
We show how performance profiling tools such as NVIDIA's Nsight Compute can be used to guide performance optimization.  
Although OpenACC is relatively easy to use, it has some limitation on resource allocation and utilization. We show that CUDA FORTRAN can provide more flexibility and allow us to fine tune the performance by managing how thread blocks and threads are configured and utilized to achieve multiple levels of concurrency as well as reducing memory allocation and host-device data transfer overhead.

The paper is organized as follows. In Section~\ref{sec:kernels}, we describe the main computational tasks in the numerical solution of the Kadanoff-Baym equation and examine different levels of concurrency in main computational kernels.  We discuss how the main computational tasks in a KBE solver can be parallelized on distributed memory many-core CPUs by using a hybrid MPI/OpenMP implementation and show that such an implementation has a good parallel scaling in Section~\ref{sec:MPI}.  In Sections~\ref{sec:ACC}, we discuss how OpenACC can be used to replace OpenMP and enable the KBE solver to run efficiently on NVIDIA GPUs. The advantage of using CUDA FORTRAN to improve the performance of the KBE solver on NVIDIA GPUs is shown in Section~\ref{sec:CUDA}. Several computational experiments are presented in Section~\ref{sec:results} to compare the performance of different programming models on CPUs and GPUs. We also show the strong and weak parallel scaling of the CPU and GPU implementation of the KBE solver in that section.

\section{Computational Kernels}\label{sec:kernels}
In this section, we will use a specific model problem to illustrate the main computational kernels used in a numerical time evolution scheme for solving the KBE \eqref{eq:kbe}. The model problem is a so-called two-band Hubbard model~\cite{hirsch1985two,white1989numerical,macridin2005physics,werner2009momentum,pudleiner2016momentum}. The periodic boundary condition for the $N_0$-site hopping term allows us to define the Hamiltonian
\begin{equation}\label{eq:Htotal}
    H_{\rm{total}}(t) = H_{\rm{s}} + H_{\rm{ext}}(t),
\end{equation}
in the momentum space. In particular, the system Hamiltonian $H_{\rm{s}}$ can be written as 
\begin{equation}
\begin{aligned}
    H_{\rm{s}}=\sum_\vk&(\epsilon_{v\vk}c^\dagger_{v\vk} c_{v\vk} + \epsilon_{c\vk}c^\dagger_{c\vk}c_{c\vk})-U(t)\sum_\vk c^\dagger_{c\vk}c_{c\vk}\\
    &+\frac{U(t)}{N}\sum_{\vk_1,\vk_2,\vq}c^\dagger_{v\vk_1+\vq}c^\dagger_{c\vk_2-\vq}c_{c\vk_2}c_{v\vk_1},
    \label{eq:ham}
\end{aligned}
\end{equation}
where $\vk$ denotes $k$-points sampled from the Brillouin zone associated with the $N$-site lattice, $c^{\dagger}_{v\vk}$ and $c_{v\vk}$ are creation and annihilation operators associated with the valence band $v$, $c^{\dagger}_{c\vk}$ and $c_{c\vk}$ are creation and annihilation operators associated with the conduction band $c$,
$\epsilon_{v\vk}$ and $\epsilon_{c\vk}$ are the band energies of the valence and conduction bands with momentum $\vk$ respectively, $U(t)$ is the time-dependent on-site interaction between the two bands.  
%The energy dispersion is given by 
%\begin{equation}
%\left\{
%\begin{aligned}
%    \epsilon_{v\vk}&=-2(1-\cos(\vk)) - E_g/2 \\
%    \epsilon_{c\vk}&=2(1-\cos(\vk)) + E_g/2
%\end{aligned}
%\right.
%\end{equation}
%with $E_g=1$ the band gap.
The external field $H_{\rm{ext}}(t)$ in \eqref{eq:Htotal} is the light-matter coupling within the dipole approximation defined by
\begin{align}
    H_{\rm{ext}}(t) = E(t)\sum_{\vk}(d_\vk c^\dagger_{c\vk}c_{v\vk}+d^*_\vk c^\dagger_{v\vk}c_{c\vk}),
    \label{eq:dipole}
\end{align}
where $E(t)=I\delta(t-0.5)$ is a time-dependent pulse centered at $t=0.5$ with intensity $I$, and $d_\vk$ is the dipole matrix element. For simplicity we set $d_\vk=1$.

The KBE associated with such a model problem can be written as
\begin{equation}\label{eq:Green}
\footnotesize
\left\{
\begin{aligned}
    &\left[i\frac{d}{dt}-h(\vk;t)\right]G_{jm}^<(\vk;t,t') = \delta(t,t') + I^<_{jm}(\vk;t,t'),\\
    &\left[-i\frac{d}{dt'}-h(\vk;t')\right]G_{jm}^>(\vk;t,t') = \delta(t,t') + I^>_{jm}(\vk;t,t'),
\end{aligned}
\right.
\end{equation} 
where the $\vk$-dependent $G^{\lessgtr}_{jm}(\vk;t,t')$ represents the lesser/greater Green's function with $j$ and $m$ being the band indices, and $I^{\lessgtr}_{jm}(\vk;t,t')$ is known as a collision integral defined as 
\begin{equation}\label{eq:CI}
\footnotesize
\begin{aligned}
I^<_{jm}(\vk;t,t') = &\int_0^{t}d\overline{t}\left[(G^>(\vk;t,\overline{t})-G^<(\vk;t,\overline{t}))\Sigma^{<}(\vk;\overline{t},t')\right]_{jm}\\
&\;+\int_0^td\overline{t}\left[G^<(\vk;t,\overline{t})(\Sigma^<(\vk;\overline{t},t')-\Sigma^>(\vk;\overline{t},t'))\right]_{jm},
\end{aligned}
\end{equation}
with a similar expression for $I^{>}_{jm}(\vk;t,t')$. 

Although the integro-differential equations \eqref{eq:Green} may appear to be completely independent for different $\vk$'s, they are actually coupled through the self-energy terms $\Sigma_{jm}^{\lessgtr}(\vk,t,t')$.

When the second-Born approximation is used to define the self-energy terms,  
we can write $\Sigma^{<}_{jm}(\mathbf{k}; t,t')$ as
\begin{equation}
\footnotesize
\begin{aligned}\label{eq:2BSE}
&\Sigma^{<}_{jm}(\mathbf{k}; t,t')\\
=&\frac{U(t)U(t')}{n_k^2}\sum_{\mathbf{q}\mathbf{k}'}G^<_{\bar{j}\bar{m}}(\mathbf{k}'+\mathbf{q}; t,t')G^>_{\bar{m}\bar{j}}(\mathbf{k}'; t',t)G^<_{jm}(\mathbf{k}-\mathbf{q}; t,t')\\
& \; -\frac{U(t)U(t')}{{n_k^2}}\sum_{\mathbf{q}\mathbf{k}'}G^<_{j\bar{m}}(\mathbf{k}'; t,t')G^>_{\bar{m}\bar{j}}(\mathbf{k}'+\vq - \mathbf{k}; t',t)G^<_{\bar{j}m}(\vq; t,t').
\end{aligned}
\end{equation}
Similar expressions can be written for $\Sigma^{>}(\vk;\overline{t},t)$. 

%In~\eqref{eq:2BSE}, we take $\bar{j}=2\,(1)$ if $j=1\,(2)$ for the band indices.
%If we propagate the NEGF through the solution to the KBEs~\eqref{eq:Green} for $n_t$ time steps, then 
%Note that dimension of the NEGF is $n_k\times n_b\times n_b\times n_t\times n_t$, with $n_k$ the number of $k$-points and $n_b=2$ the number of bands. 

The numerical method we use to solve~\eqref{eq:Green} advances $G^\lessgtr(t,t')$ along both the $t$ and the $t'$ directions. If we let $(t_i,t'_j) \equiv (i\Delta t,j\Delta t)$, be a grid point on a uniform two-time grid, the numerical propagation of $G^<$ according to the first equation in~\eqref{eq:kbe} is described in Algorithm~\ref{alg:prop}.

\begin{algorithm}[ht!]
\small
\caption{The propagation of $G^{<}$ at the $n_t$-th time step}\label{alg:prop}
\begin{algorithmic}[1]
\Require{$G^\lessgtr(t_j,t'_{\ell})$, $j,\ell=0,...,n_t-1$, $max$, $\varepsilon$}
\Ensure{$G^<(t_{n_t}, t'_\ell)$, $\ell=0,...,n_t$}

\For{$\ell\gets 0$ to $n_t$}
\State Evaluate $h(t_{n_t-1} + \Delta t/2)$
\State Evaluate $\Sigma^\lessgtr(t_j,t'_{n_t-1})$ and $\Sigma^\lessgtr(t_{n_t-1},t'_\ell)$ by~\eqref{eq:2BSE}
\State Evaluate $I^<(t_{n_t-1},t'_\ell)$ by~\eqref{eq:CI}
\State Solve~\eqref{eq:Green} to get $G^<(t_{n_t}, t'_{\ell})$
\EndFor
\For{$i\gets 1$ to $max$}
\For{$\ell\gets 0$ to $n_t$}
\State $G^<_0(t_{n_t}, t'_\ell)\gets G^<(t_{n_t}, t'_\ell)$
\State Evaluate $h(t_{n_t})$
\State Evaluate $\Sigma^\lessgtr(t_j,t_{n_t})$ and $\Sigma^\lessgtr(t_{n_t},t'_\ell)$ by~\eqref{eq:2BSE}
%\JY{here we need to use $G^>(t_j, t'_{n_t})$}
\State Evaluate $I^<(t_{n_t},t'_\ell)$ by~\eqref{eq:CI}
\State $I_1^<(t_{n_t-1},t'_\ell)\gets \frac{1}{2}\left(I^<(t_{n_t-1},t'_\ell)+I^<(t_{n_t},t'_\ell)\right)$ 
\State Solve~\eqref{eq:Green} with $I_1^<$ to get $G^<(t_{n_t}, t'_\ell)$ 
\EndFor
\If{$\|G^<(t_{n_t}, \cdot)-G_0^<(t_{n_t}, \cdot)\|\leq\varepsilon$}
  \State \textbf{break}
\EndIf
\EndFor
\end{algorithmic}
\end{algorithm}

\begin{figure}[ht!]
    \centering
    \includegraphics[width=0.9\linewidth]{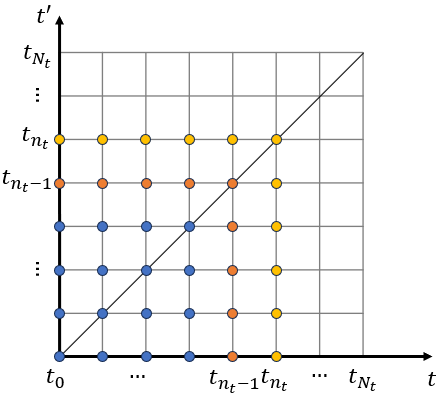}
    \caption{Illustration of the evaluation at the $n_t$-th time step for $\Sigma^\lessgtr$ and $I^\lessgtr$. The blue dots represent values computed at previous time steps, and the orange dots give the $2n_t-1$ pairs of $(t,t')$ at which the values of $\Sigma^\lessgtr$ and $I^\lessgtr$ are computed at the $n_t$-th time step.}
    \label{fig:evolv}
\end{figure}

At the $n_t$-th time step, when advancing from $G^{<}(t_{n_t-1},t'_\ell)$ to $G^<(t_{n_t},t'_\ell)$, $\ell=0, .., n_t-1$ we need to 
\begin{itemize}
\item Evaluate the single-particle Hamiltonian $h(t_{n_{t_1}}+\Delta t/2$);
\item Evaluate the self-energies $\Sigma^{\lessgtr}(t_j,t'_{n_t-1})$ and
$\Sigma^{\lessgtr}(t_{n_t-1}, t'_\ell)$ defined by~\eqref{eq:2BSE} with $j,\ell=0,...n_t-1$, i.e., the self energies on the orange time grid points in Fig.~\ref{fig:evolv};
\item Evaluate the collision integral $I^{<}(t_{n_t-1},t'_\ell)$ defined by~\eqref{eq:CI}, $\ell=0, ..., n_t-1$, i.e. the collision integrals along the $n_t$th column of the two time grid points (marked in orange in Fig.~\ref{fig:evolv};
\item Numerically evolve $G^{<}(t_{n_t-1},t_\ell)$ to $G^{<}(t_{n_t},t_\ell)$, $\ell=0,...,n_t-1$ using an appropriate time stepping scheme~\cite{balzer2007nonequilibrium,Stan2009,bonitz1996numerical,hoppensteadt2007numerical,kennedy2016diagonally}. Because \eqref{eq:Green} is nonlinear, a non-linear iteration is often performed to ensure the consistency of $G$ and the right-hand side of \eqref{eq:Green}. To reach self-consistency, we need to repeatedly evaluate the single-particle Hamiltonian $h(t_{n_t})$, the self-energies $\Sigma^\lessgtr(t_j,t'_{n_t})$, $\Sigma^\lessgtr(t_{n_t}, t'_\ell)$ and the collision integral $I^{<}(t_{n_t},t'_\ell)$ with $j,\ell=0,...n_t$, i.e., self-energies and collision integrals along the yellow time grid points marked in Fig.~\ref{fig:evolv}.
\end{itemize}
A similar set of operations are performed to advance $G^{>}(t_j,t_{n_t-1})$ to $G^{>}(t_j,t_{n_t})$ with $j=0,...,n_t$. 

Note that the self-energy $\Sigma^{<}$ defined in \eqref{eq:2BSE} is the difference of two terms. We denote the first and second terms by $\Sigma^1_{jm}(\mathbf{k}; t,t')$ and $\Sigma^2_{jm}(\mathbf{k}; t,t')$ respectively.

We compute $\Sigma^1_{jm}(\mathbf{k}; t,t')$ in two steps. In the first step, we compute the polarizability $P_{jm}(\vq; t, t')$
\begin{equation}
\small
    P_{jm}(\vq; t, t') = \sum_{\vk'}G^<_{{j}{m}}(\mathbf{k'+q}; t,t')G^>_{{m}{j}}(\mathbf{k'}; t',t), \label{eq:polarizability}
\end{equation}
This step contracts $G^<_{jm}$ and $G^>_{mj}$ over $\vk'$. The complexity for such a contraction is $\mathcal{O}(n_k)$. Since we need to perform the contraction for all $k$-points, for given $j$, $m$, $t$, $t'$, the complexity to compute $P_{jm}(\vq; t, t')$ is $\mathcal{O}(n_k^2)$.

In the second step, we contract $P_{\bar{j}\bar{m}}$ obtained above with $G^<_{jm}$ over $\vq$ to get  $\Sigma^1_{jm}(\mathbf{k}; t,t')$
\begin{equation}
\small
    \Sigma^1_{jm}(\mathbf{k}; t,t') =  U(t)U(t')\sum_{\vq}P_{\bar{j}\bar{m}}(\vq; t, t')G^<_{jm}(\mathbf{k-q}; t,t').\label{eq:kernel2}
\end{equation}
This contraction also costs $\mathcal{O}(n_k)$ for one $k$-point, and $\mathcal{O}(n_k^2)$ considering all $k$-points. 
%This two-step algorithm is demonstrated in Algorithm~\ref{alg:contraction1}. In the algorithm, we omit the time arguments $t$ and $t'$ for brevity.

By breaking the contractions over $\vk'$ and $\vq$ required in \eqref{eq:2BSE} into two steps, we reduce the $k$-point complexity from $\mathcal{O}(n_k^3)$ to $\mathcal{O}(n_k^2)$. 
%\begin{algorithm}[ht!]
%\small
%\caption{The pseudocode for computing $\Sigma^1_{jm}(\mathbf{k})$}\label{alg:contraction1}
%\begin{algorithmic}[1]
%\Require{$j$, $m$, $G^\lessgtr$}
%\Ensure{$\Sigma^1_{jm}(\vk)$}
%\State $\Sigma^1_{jm}(\vk)\gets 0$
%\For{$q\gets 1$ to $n_k$}
%  \State $P(q)\gets 0$
%  \For{$k'\gets 1$ to $n_k$}
%    \State $P(q)\gets P(q)+G^<_{\bar{j}\bar{m}}(k'+q)G^>_{\bar{m}\bar{j}}(k')$
%  \EndFor
%\EndFor
%\For{$k\gets 1$ to $n_k$}
%  \For{$q\gets 1$ to $n_k$}
%    \State $\Sigma^1_{jm}(k)\gets \Sigma^1_{jm}(k)+P(q)G^<_{jm}(k-q)$
%  \EndFor
%\EndFor
%\end{algorithmic}
%\end{algorithm}

However, such a reduction in complexity is not applicable to $\Sigma^2_{jm}(\mathbf{k}; t,t')$, i.e., the contractions over $\vk'$ and $\vq$ cannot be decoupled. As a result, the evaluation of $\Sigma^2_{jm}(\mathbf{k}; t,t')$ takes $\mathcal{O}(n_k^3)$ operations. The implementation of such a calculation follows the nested loops  shown in the pseudocode given in Algorithm~\ref{alg:contraction2}. We omit the time arguments $t$ and $t'$ in the pseudocode for clarity.
\begin{algorithm}[ht!]
\small
\caption{The evaluation of $\Sigma^2(\mathbf{k})$}\label{alg:contraction2} \begin{algorithmic}[1]
%Require{$j$, $m$, $G^\lessgtr$}
%\Ensure{$\Sigma^2_{jm}(\vk)$}
\State $\Sigma^2(\vk)\gets 0$
\For{$k\gets 1$ to $n_k$}
  \For{$j\gets 1$ to $2$}
    \For{$m\gets 1$ to $2$}
      \For{$q\gets 1$ to $n_k$}
        \For{$k'\gets 1$ to $n_k$}
          \State $\Sigma^2_{jm}(k)\gets \Sigma^2_{jm}(k)+$\par  \hskip2.5cm$G^<_{j\bar{m}}(k')G^>_{\bar{m}\bar{j}}(k'+q-k)G^<_{\bar{j}m}(q)$
        \EndFor
      \EndFor
    \EndFor
  \EndFor
\EndFor
\end{algorithmic}
\end{algorithm}

%and we will mainly focus on optimizing it in the following discussions for the self-energy computation.

%The main loop to compute the collision integral in the \texttt{nt}-th time step is similar to Fig.~\ref{fig:workflow}. Instead of calling \texttt{get\_2B\_SE}, we call the subroutine \texttt{get\_CI}. 
The collision integral~\eqref{eq:CI} can be evaluated as the sum of two terms $I^1_{jm}(\vk;t,t')$ and $I^2_{jm}(\vk;t,t')$ respectively.

Since the evaluations of $I^1$ and $I^2$ are similar once we have $G^\lessgtr$ and $\Sigma^\lessgtr$ computed and stored, we will only discuss how $I^2$ is computed below.

The main computation performed in the evaluation of $I^2$ is the convolution of $G^<(\vk,t,\bar{t})$ and $\Sigma^<(\vk,\bar{t},t')-\Sigma^>(\vk,\bar{t},t')$. Using a standard second order numerical integration scheme to compute such a convolution requires $\mathcal{O}(n_t)$ floating point operations at the $n_t$-th time step for each $k$-point.
%matrix differences, matrix products, and numerical integration as demonstrated in Algorithm~\ref{alg:CI} for the $n_t$-th time step. From the pseudocode, it is clear that the complexity is $\mathcal{O}(n_kn_t)$.
%\begin{algorithm}[ht!]
%\small
%\caption{The pseudocode for computing $I^2(\mathbf{k})$}\label{alg:CI}
%\begin{algorithmic}[1]
%\Require{$G^<$, $\Sigma^\lessgtr$}
%\Ensure{$I^2(\mathbf{k})$}
%\For{$s\gets 1$ to $n_t$}
%  \For{$k\gets 1$ to $n_k$}
%    \State $\Sigma^R(k;s)\gets \Sigma^<(k;s)-\Sigma^>(k;s)$
%  \EndFor
%\EndFor
%\For{$k\gets 1$ to $n_k$}
%  \For{$s\gets 1$ to $n_t$}
%    \State $tmp(s) = G^<(k;s)\Sigma^R(k;s)$ 
%  \EndFor
%  \State $I^2(k)\gets$ integration of $tmp(s)$ over $s$ from $1$ to $n_k$
%\EndFor
%\end{algorithmic}
%\end{algorithm}

Algorithm~\ref{alg:prop} shows that we need to perform at least $n_t$ self-energy and collision integral calculation in order to obtain approximations to $G^{\lessgtr}(t_j,t'_{n_t})$ and $G^{\lessgtr}(t_{n_t},t'_j)$, for $j=1,2,...,n_t$. As a result,
the overall complexities for the self-energy and the collision integral computation become $\mathcal{O}(n_k^3n_t)$ and $\mathcal{O}(n_kn_t^2)$ respectively. Once these quantities become available, evolving $G^\lessgtr$ from $(t_j,t'_{n_t-1})$ to $(t_j,t'_{n_t})$ and from $(t_{n_t-1},t'_j)$ to $(t_{n_t},t'_{j})$ 
takes only $\mathcal{O}(n_kn_t)$ operations, which is considerably lower. Therefore, when $n_t$ is small, the time stepping cost is dominated by the self-energy computation. The collision integral evaluation only becomes more expensive when $n_t$ becomes much larger than $n_k$. If we evolve the equation for a large $N_t$ time steps, the overall computational cost will be $\mathcal{O}(n_k^3N_t^3)$. This cost increases rapidly with increasing $n_k$ or $N_t$. A simple implementation shows that for $n_k=1024$, performing $10$ time steps sequentially takes more than 1 day to compute on a modern CPU, which is prohibitively expensive for solving problems of interest.

%Since in the time evolution, each time step relies on the results from previous steps, we have to evolve the Green's function sequentially. \CY{contradict earlier statement about time parallelism}\CY{what dependency?} Consequently, we will focus on parallelizing the computation at a single time step. As in the $j$-th time step, $\Sigma^{\rceil/\lceil}$ is only updated for one $t$ entry compared to $j$ entries for $\Sigma^{\lessgtr}$, our implementation center on the computation for $\Sigma^{\lessgtr}(\vk;t,t')$.

\section{MPI/OpenMP implementation}\label{sec:MPI}
The separation of the Kadanoff-Baym equation \eqref{eq:Green} by $k$-points for the two-band Hubbard model as shown in \eqref{eq:2BSE} naturally suggests that we can distribute the time evolution of $G_{jm}^{\lessgtr}(\vk,t,t')$ among different MPI ranks by $k$-points. 
Note that the single-particle Hamiltonian $h$ is replicated on all MPI ranks, the collision integral $I^{\lessgtr}$, which is distributed among different MPI ranks by $k$-points, and the update of $G^{\lessgtr}$ can all be computed independently for different $\vk$'s. 
The evaluations of the self-energy terms $\Sigma^{\lessgtr}$, which require contractions of $G^{\lessgtr}$ over all $k$-points, can be distributed by $k$-points also. In particular, we can distribute the outermost loop in Algorithm~\ref{alg:contraction2} among $n_p$ MPI ranks, with each MPI rank loop over $\mathrm{my\_nk} = n_k/n_p$ $k$-points (assuming $n_p$ divides $n_k$.) 
Collective communications ({\tt{MPI\_Allreduce}}) are required to replicate the Green's functions on all MPI ranks so that we can use the complete Green's function to compute distributed (local) self-energy tensor simultaneously. The main steps of the KBE solver are shown in Fig.~\ref{fig:workflow}.
In each time step, we evaluate the single particle Hamiltonian by calling the subroutine {\tt{get\_h}}. This is followed by calling self-energy and collision integral evaluation subroutines {\tt{get\_SE\_at\_t}} and {\tt{get\_CI\_at\_t}}. A time evolution method is then used in {\tt{time\_step}} to advance the Green's function to the next time slices on the two-time grid.

\begin{figure}[ht!]
\centering
\begin{minipage}{1\linewidth}
    \footnotesize
    \begin{tabular}{c}
    \begin{lstlisting}
do it=1,nt
  ! compute the single-particle Hamiltonian
  call get_h(ham,...)
  ! compute the self-energy
  call get_SE_at_t(se,...)
  ! compute the collision integral
  call get_CI_at_t(ci,...)
  ! evolve the Green's function
  call time_step(Gg,Gl,...)
  ! MPI reduce
  call MPI_Allreduce(Gg,gt(...,it+1,:it),size(Gg),&
             MPI_COMPLEX_DPC,MPI_SUM,...)
  call MPI_Allreduce(Gl,gt(...,:it+1,it+1),size(Gl),&
             MPI_COMPLEX_DPC,MPI_SUM,...)
enddo
    \end{lstlisting}
    \end{tabular}
    \caption{The main steps in the outermost loop of the KBE solver.}
    \label{fig:workflow}
\end{minipage}
\end{figure}

We use CrayPat to analyze the performance of the MPI implementation.
In a run that propagates the Green's function for 10 time steps on 1024 MPI ranks with $n_k = 1024$ k-points, the profiler shows that the self-energy computation takes over $95\%$ of the total time.  This observation is consistent with our previous cost analysis that suggests self energy evaluate would dominate the entire computation when $n_t \ll n_k$.
Therefore, in the following, we focus first on investigating the performance of the self-energy computation. 

In Fig.~\ref{fig:MPI}, we show a code snippet for computing $\Sigma_{jm}^2(\vk,t,t')$ at a fixed $(t,t')$ pair in parallel among $n_p$ MPI ranks. We recall that the Green's functions are updated locally and stored globally. Arrays \texttt{Gtmp1} and \texttt{Gtmp2} in the code snippet store the Green's functions at all $n_k$ $k$-points.

\begin{figure}[ht!]
\centering
\begin{minipage}{0.97\linewidth}
\footnotesize
\begin{tabular}{c}
\begin{lstlisting}
! loop over local k-points 
do kk=1,my_nk
  ! map local k-point indices to global indices
  ikk = bz%my_kpts(kk) 
  do jj=1,dim
    do ii=1,dim    
      do kp=1,nk
        do kpp=1,nk
          iqk = bz%allind_kpk(kp,kpp)
          iqk = bz%allind_kmk(iqk,ikk)
          se2(kk,ii,jj) = se2(kk,ii,jj)+Ut*Utp*&
                  Gtmp1(kp,ii,dim-jj+1)*&
                  Gtmp2(iqk,dim-jj+1,dim-ii+1)*&
                  Gtmp1(kpp,dim-ii+1,jj)
        enddo
      enddo
    enddo
  enddo
enddo
\end{lstlisting}
\end{tabular}

\end{minipage}
\caption{The MPI implementation of $\Sigma^2_{jm}(\vk;t,t')$ computation at a given $(t,t')$ pair.}
\label{fig:MPI}
\end{figure}

The nested loop structure shown in Algorithm~\ref{alg:contraction2} and Fig.~\ref{fig:MPI} indicates that within each MPI rank, additional levels of concurrency for the self-energy evaluation can be exploited by using OpenMP.

Because the number of bands (2) in the model Hamiltonian is far fewer than the number of threads within each MPI rank (on a manycore compute node) compared to the number of $k$-points, we reorder the nested loops in Fig.~\ref{fig:MPI} to move the {\tt{kp}} and {\tt{kpp}} loops out and keep the {\tt{jj}} and {\tt{ii}} loops as the inner loops. An {\tt{!OMP PARALLEL DO}} directive is placed right before the {\tt{kp}} loop to execute the outer loop in parallel. A local private temporary array {\tt{se2tmp}} is created to accumulate the intermediate contraction results. A reduction is performed through the use of the directive {\tt{REDUCTION(+:se2tmp)}}to sum up contributions from all threads. The hybrid MPI/OpenMP implementation of the $\Sigma_{jm}^2(\vk,t,t')$ calculation is shown in Fig.~\ref{fig:OMP}.

\begin{figure}[ht!]
\centering
\begin{minipage}{0.98\linewidth}
\footnotesize
\begin{tabular}{c}
\begin{lstlisting}
do kk=1,my_nk
  ikk = bz%my_kpts(kk) 
  !$OMP PARALLEL DO PRIVATE(iqk) REDUCTION(+:se2tmp)
  do kp=1,nk
    do kpp=1,nk
      iqk = bz%allind_kpk(kp,kpp)
      iqk = bz%allind_kmk(iqk,ikk)
      do jj=1,dim
        do ii=1,dim
          se2tmp(ii,jj,kk) = se2tmp(ii,jj,kk)+&
                  Ut*Utp*Gtmp1(kp,ii,dim-jj+1)*&
                  Gtmp2(iqk,dim-jj+1,dim-ii+1)*&
                  Gtmp1(kpp,dim-ii+1,jj)
        enddo
      enddo
    enddo
  enddo
  !$OMP END PARALLEL DO
  se2(kk,:,:) = se2tmp(:,:,kk)
enddo
\end{lstlisting}
\end{tabular}
\end{minipage}
\caption{The hybrid MPI/OPenMP implementation of the $\Sigma^2_{jm}(\mathbf{k}; t,t')$ computation at a given $(t,t')$ pair.}
\label{fig:OMP}
\end{figure}

\begin{figure}[t]
    \centering
    \includegraphics[width=0.9\linewidth]{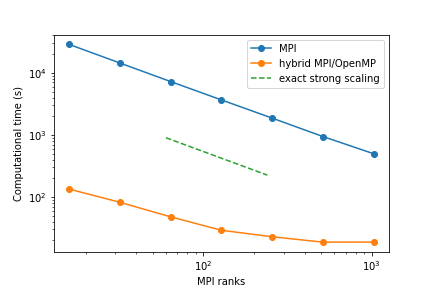}
    \caption{The performance of the MPI and the hybrid MPI/OpenMP implementation to the program with varying number of MPI ranks. In the test problem, $n_k=1024$, and we compute the first $10$ time steps.}
    \label{fig:MPI_perf}
\end{figure}

Fig.~\ref{fig:MPI_perf} shows how the non-equilibrium Green's function computation scales with respect to the number of MPI ranks.
This calculation is performed on the Perlmutter CPU nodes maintained at National Energy Research Scientific Computing (NERSC) Center.
Each CPU node is composed of 2 AMD EPYC 7763 (Milan) CPUs with 64
cores per CPU, and 2 threads per core. In total, there is 512
GB DDR4 memory for each node, and for each CPU, the memory bandwidth is 204.8 GB/s.  In this experiment, we set the number of k-points to $n_k=1024$, the timing measurement  for the self-energy and the collision integral calculation is collected for the first $10$ time steps.

We observe that the wall clock time used decreases almost linearly with respect to the number of MPI ranks, which indicates that the overhead incurred in {\tt{MPI\_Allreduce}} is relatively small. 

In Fig.~\ref{fig:MPI_perf}, we also show how the hybrid MPI/OpenMP version of the program scales with respect to MPI ranks. In these calculations, we use $64$ threads within each MPI rank, and each MPI rank is bind to a compute node. 
We observe a nearly perfect strong scaling when the number of MPI ranks is less than $128$. Increasing the number of MPI ranks beyond 128 results in suboptimal strong scaling due to  the OpenMP reduction overhead relative to the amount of work per MPI rank, which decreases as the number of MPI ranks increases.

\section{GPU Implementation with OpenACC}\label{sec:ACC}

Although the MPI implementation exhibits nearly perfect strong scaling, Fig.~\ref{fig:MPI_perf} shows that wall clock time required to 
evolve the Green's function for $10$ time steps already exceeds $8$ minutes even with $1024$ MPI ranks. As mentioned before, the calculation of the self-energy and the collision integral takes over $95\%$ of the total computational time.  The use of hybrid MPI/OpenMP parallelization allows us to reduce the computation time by a factor of 26 on $1024$ MPI ranks with $64$ threads per MPI rank. However, when the number of time steps increases, the self-energy and collision integral evaluation will become progressively more time-consuming. To obtain a trajectory with $N_t=500$ time steps, the hybrid MPI/OpenMP implementation uses more than $3$ hours to compute the self-energy and about $2$ hours to evaluate the collision integral.

To further improve the throughput of NEGF computation, we investigated the possibility of running the KBE solver efficiently on high performance computers equipped with accelerators. In this work, we focus on enabling the KBE solver to run efficiently on NVIDIA graphics processing units (GPUs).

%\CY{not sure if this is the right place for this}
%To further accelerate the computation, we aim to make full use of the computing power of GPU to introduce more parallelism in both space and time \CY{do we have parallelism in time?}\JY{Not for the time evolution, but we use batching for time in a single time step} through OpenACC.
%Our development is based on a Fortran code applying the implicit second-order Runge-Kutta method~\cite{Stan2009} to propagate the NEGF.

Compared to CPUs, GPUs have a large number of streaming multiprocessors (SMs) with multiple cores each and thousands of threads organized into thread blocks and warps.  The Single Instruction, Multiple Thread (SIMT) execution model and the availability of a large amount of high bandwdith shared memory and local cache makes GPUs ideal for computations that have high arithmetic intensity and regular memory access patterns.

To exploit the tremendous amount of computational resources available on GPUs, we need to design the solver to map concurrent tasks performed in the KBE solver within an MPI rank to different threads or thread blocks excecuted simultaneously on a GPU device.  On NVIDIA GPUs, the easiest programming model that allows us to achieve such a mapping is the OpenACC directive-based programming~\cite{ACC,chandrasekaran2017openacc}, which is an open source parallel programming standard. 

Although a GPU node has much higher peak performance than a CPU node, there exist a few challenges that can hinder the optimal execution of code on a GPU. In particular, repeated memory allocations on the GPU device and excessive data movement between the host CPU and the GPU device can introduce a significant amount of overhead. The execution of a computational task on the GPU requires a kernel launch. A large number of repeated kernel launches can also degrade the performance of the overall computation. Furthermore, a sub-optimal memory access pattern that results in a memory bank conflict can also contribute to sub-optimal performance of the code.

In this section, we describe how to overcome these challenges when OpenACC directives are used to offload the $\Sigma^2$ and $I^2$ calculations onto NVIDIA GPUs. 
We can use OpenACC directives to manage memory access and map computational work to multiple thread blocks and threads to achieve optimal utilization of GPU resources.

%To further utilize the high concurrency in computing $\Sigma^2_{jm}(\vk;t,t')$ shown in Algorithm~\ref{alg:contraction2}, we need to optimize the block and thread configurations by using the OpenACC directives.

We will demonstrate various performance improvement resulting from the use of OpenACC directives through a test problem that uses $n_k=1024$, $n_t=500$. The problem is solved on $4$ MPI ranks.
%\JY{\texttt{my\_nk} is set to $4$ manually here to quickly check the results. This is the normal \texttt{my\_nk} in actual problems.}

\subsection{Exploiting GPU concurrency by loop fusion and reordering}
OpenACC provides a \texttt{parallel loop} directive similar to the \texttt{parallel do} directive in OpenMP. Therefore, we can simply replace the OpenMP directive in Fig.~\ref{fig:OMP} with the OpenACC directive to enable the self-energy calculation to be performed on a GPU.  However, because the number of threads on a GPU far exceeds that on a CPU, one cannot effectively utilize all threads if \texttt{nk} is not sufficiently large.  To address this issue, we use the \texttt{collapse} clause in OpenACC to fuse the \texttt{kp} and \texttt{kpp} loops as shown in Fig.~\ref{fig:accloopfusion} to create more concurrent work that can be mapped to a large number of threads and executed simultaneously. A reduction operation is performed to sum up contributions from different threads into a scalar temporary variable \texttt{tmp} because OpenACC does not support reduction onto an element of an array directly. This limitation also requires us to move the {\texttt{jj}} and {\texttt{ii}} loops outside of the \texttt{kp} and \texttt{kpp} loops so that reduction is performed over $\texttt{kp}$ and $\texttt{kpp}$ for each (\texttt{ii}, \texttt{jj}) pair. 
Such a simple modification significantly improves the performance of the self-energy calculation on a GPU and reduces the wallclock time used to complete this part of calculation for the test problem by $86.78\%$ compared to the baseline implementation without using \texttt{collapse}.

\begin{figure}[ht!]
\centering
\begin{minipage}{0.93\linewidth}
\footnotesize
\begin{tabular}{c}
\begin{lstlisting}
do kk=1,my_nk
  ikk = bz%my_kpts(kk)
  do jj=1,dim
    do ii=1,dim
      tmp = (0d0,0d0)
!$acc parallel loop reduction(+:tmp) collapse(2)
      do kp=1,nk
        do kpp=1,nk
          iqk = bz%allind_kpk(kp,kpp)
          iqk = bz%allind_kmk(iqk,ikk)
          tmp = tmp+Ut*Utp*Gtmp1(kp,ii,dim-jj+1)&
              * Gtmp2(iqk,dim-jj+1,dim-ii+1)&
              * Gtmp1(kpp,dim-ii+1,jj)
        enddo
      enddo
!$acc end parallel loop
      se2(kk,ii,jj) = tmp
    enddo
  enddo
enddo
\end{lstlisting}
\end{tabular}
\end{minipage}
\caption{Using OpenACC \texttt{parallel loop} directive and loop fusion to accelerate $\Sigma^2_{jm}(\mathbf{k}; t,t')$ computation for a $(t,t')$ pair on a GPU.}
\label{fig:accloopfusion}
\end{figure}

Another way to leverage the GPU architecture is to exploit multiple levels of concurrency that exists in the nested loops in Algorithm~\ref{alg:contraction2}. This can be achieved by using the \texttt{loop gang} and \texttt{loop vector} directives provided in OpenACC to leverage the thread block, warp and thread hierarchy on NVIDIA GPUs. 
 %we need to reorder the nested loops in, for example, the self-energy calculation, to maximize thread utilization and achieve a high level of concurrency. 
 The code snippet displayed in Fig.~\ref{fig:ACC_unbatch} shows where the \texttt{loop gang} and \texttt{loop vector} directives are placed. 
% In order to enable more effective SIMT execution of the inner loop, and place the \texttt{loop vector} directive before the \texttt{kp} loop to allow multiple threads to execute different \texttt{kp} iterates simultaneously. 
Both the outer loops (at \texttt{kk}, \texttt{jj} and \texttt{ii} levels) and the inner loops (at \texttt{kp} and \texttt{kpp} levels) are automatically fused by the compiler to maximize thread block and thread utilization. This modification results in $85.32\%$ performance improve compared to the baseline implementation.  The slightly smaller performance improvement compared to simply using the \texttt{collapse} directive to fuse the loops over \texttt{kp} and \texttt{kpp} is likely due to the small range of \texttt{my\_nk}, \texttt{jj} and \texttt{ii} values that limit the number of thread blocks that can be fully utilized.

\begin{figure}[ht!]
\centering
\begin{minipage}{0.97\linewidth}
\footnotesize
\begin{tabular}{c}
\begin{lstlisting}
!$acc kernels loop gang present(se2)
do kk=1,my_nk
  do jj=1,dim
    do ii=1,dim
      ikk = bz%my_kpts(kk)
      tmp = (0d0,0d0)  
      !$acc loop vector reduction(+:tmp)
      do kp=1,nk
        do kpp=1,nk
          iqk = bz%allind_kpk(kp,kpp)
          iqk = bz%allind_kmk(iqk,ikk)
          tmp = tmp+Ut*Utp*Gtmp1(kp,ii,dim-jj+1)*&
                Gtmp2(iqk,dim-jj+1,dim-ii+1)*&
                Gtmp1(kpp,dim-ii+1,jj)
        enddo
      enddo
      !$acc end loop
      se2(kk,ii,jj) = tmp
    enddo
  enddo
enddo
!$acc end kernels
\end{lstlisting}
\end{tabular}
\end{minipage}
\caption{The OpenACC implementation for computing the  $\Sigma^2_{jm}(\mathbf{k}; t,t')$ (saved in \texttt{se2}) at a given $(t, t')$ pair. Here the \texttt{ikk} assignment is placed inside \texttt{ii} loop to ensure the \texttt{kk}, \texttt{jj} and \texttt{ii} loops can be fused.}
\label{fig:ACC_unbatch}
\end{figure}

\subsection{Updating self-energy using a single kernel launch}
In a naive implementation of Algorithm~\ref{alg:prop} that loops over the $j$ and $\ell$ indices to update $\Sigma^{<}(t_j,t'_{n_t-1})$ and $\Sigma^{>}(t_{n_t-1},t'_{\ell})$, multiple kernel launches are issued as we execute the code shown in Fig.~\ref{fig:ACC_unbatch}, which is contained in a subroutine named
\texttt{get\_2B\_SE}.  
Fig.~\ref{fig:workflow_nt} shows how \texttt{get\_2B\_SE} is called repeatedly in an outer loop over $t_l$ and $t_k'$.

Because each \texttt{itp} iterate in Fig~\ref{fig:workflow_nt} is completely independent from each other, we can exploit this level of concurrency by moving the \texttt{do itp=1,nt-1} loop inside the \texttt{!\$acc kernels loop gang} directive within the  \texttt{get\_2B\_SE} subroutine which we rename to \texttt{get\_2B\_SE\_kernel}.
A snippet of the restructured subroutine is shown in Fig.~\ref{fig:ACC}. In this subroutine, the \texttt{kernels loop gang} is placed right before the outermost \texttt{it} loop to launch the kernel once. Several temporary arrays such as \texttt{Gtmp1} and \texttt{Gtmp2} as well as input arguements such as \texttt{Ut} and \texttt{Utp} are copied from the host to the device before the kernel is launched. Once the self-energy calculation is completed on the device, the result is updated on the host at the end of the subroutine.

Fig.~\ref{fig:workflow_batch} shows how this subroutine is called. Note that instead of passing a single time snapshot of the \texttt{se} and \texttt{G} arrays, we pass the entire time slice of these arrays in either $t$ or $t'$ up to \texttt{nt-1} time steps. These time slices are shown as the red dots in Fig.~\ref{fig:evolv}. 

%By performing the kernel launch before the iterate \texttt{it} in this restructured implementation within the subroutine \texttt{get\_2B\_SE\_batch} shown in Fig.~\ref{fig:ACC}, we increase the number of gangs. This enhancement results in improved concurrency at the gang level.

\begin{figure}[ht!]
\centering
\begin{minipage}{1\linewidth}
    \footnotesize
    \begin{tabular}{c}
    \begin{lstlisting}
do itp=1,nt-1
  ! compute SE_lesser
  call get_2B_SE(se(:,:,:,itp,nt), bz, &
          G(:,:,:,itp,nt), G(:,:,:,nt,itp), Ut, Utp)
  ! compute SE_greater
  call get_2B_SE(se(:,:,:,nt,itp), bz, &
          G(:,:,:,nt,itp), G(:,:,:,itp,nt), Utp, Ut)
enddo 
    \end{lstlisting}
    \end{tabular}
    \caption{The main loop for computing the self-energy.}
    \label{fig:workflow_nt}
\end{minipage}
\end{figure}
%The variable \texttt{tbatch} in the code snippet shown in Fig.~\ref{fig:ACC} defines the range of $t$'s contained in each batch. 
%As a result, the total number of kernel launches is reduced from $n_t$ to $\lceil n_t/\texttt{batch\_size}\rceil$. we move the outer loop \ref{fig:workflow_nt}

%Because each kernel launch can introduce a noticeable amount of latency and other types of overhead, repeated kernel launches can be costly especially when the number of time steps ($n_t$) required to evolve the solution of the KBE is large.
%In addition, as we will discuss more in Section~\ref{sec:ACC1}, in the self-energy computation, we need to update several temporary arrays either on a CPU or on a GPU in each call to \texttt{get\_2B\_SE}. This update can result in additional overhead due to synchronization and data movement cost. Similar issues arise in the evaluation of the collision integral.

A similar code restructuring is used to implement a  version of  collision integral computation that uses a single kernel launch. We omit the corresponding code snippets here for brevity.
%\texttt{num\_iter} = \texttt{floor}((\texttt{nt-1})/\texttt{batch\_size}) gives the number of iterations over $t$ or $t'$ after batching. \texttt{nt\_last} = \texttt{nt-num\_iter*batch\_size} gives the batch size after the iteration when $n_t$ is not a multiple of \texttt{batch\_size}. If \texttt{nt\_last} = \texttt{1}, then we need to call the unbatched version for the last step instead. 
This approach allows us to achieve high concurrency and considerably reduces the amount of time spent in the self-energy and collision integral calculations. For example,
using this restructured code results in a $90.17\%$ performance improvement in the self-energy calculation compared with the version that implements only the loop reordering shown in Fig.~\ref{fig:ACC_unbatch} for the model problem described in Section~\ref{sec:kernels}.

\begin{figure}[ht!]
\centering
\begin{minipage}{0.97\linewidth}
\footnotesize
\begin{tabular}{c}
\begin{lstlisting}
! initialize P,se1,se2 to 0
...
!$acc data copyin(Gtmp1, Gtmp2, Ut, Utp)
!$acc update device(P,se1,se2)

!$acc kernels loop present(P)
...
!$acc end kernels

!$acc kernels loop present(P,se1)
...
!$acc end kernels

!$acc kernels loop gang present(se2)
do it=1,nt-1
  do jj=1,dim
    do ii=1,dim 
      do kk=1,my_nk
        ikk = start_kpts+kk-1
        tmp = (0d0,0d0)
        !$acc loop vector reduction(+:tmp)
        do kp=1,nk
          do kpp=1,nk
            ! compute the k-point index on the fly
            ...       
            tmp = tmp+Ut(it)*Utp(it)*&
                  Gtmp1(kp,ii,dim-jj+1,it)*&
                  Gtmp2(iqk,dim-jj+1,dim-ii+1,it)*&
                  Gtmp1(kpp,dim-ii+1,jj,it)
          enddo
        enddo
        !$acc end loop
        se2(kk,ii,jj,it) = tmp
      enddo
    enddo
  enddo
enddo
!$acc end kernels

!$acc update self(se1,se2)
!$acc end data
\end{lstlisting}
\end{tabular}
\end{minipage}
\caption{Kernel launches inside the subroutine \texttt{get\_2B\_SE\_kernel} which contains an OpenACC implementation of the self-energy computation at all time locations with a single kernel for each step.}
\label{fig:ACC}
\end{figure}
\begin{figure}[ht!]
\centering
\begin{minipage}{0.97\linewidth}
   \footnotesize
   \begin{tabular}{c}
   \begin{lstlisting}
!$acc enter data copyin(P,se1,se2)
call get_2B_SE_kernel(se(:,:,:,1:nt-1,nt), my_nk,&
      G(:,:,:,1:nt-1,nt), G(:,:,:,nt,1:nt-1),&
      U_t2(1:nt-1), U_t1(1:nt-1), P, se1, se2)
call get_2B_SE_kernel(se(:,:,:,nt,1:nt-1), my_nk,&
      G(:,:,:,nt,1:nt-1), G(:,:,:,1:nt-1,nt),&
      U_t1(1:nt-1), U_t2(1:nt-1), P, se1, se2)
!$acc exit data delete(P,se1,se2)
   \end{lstlisting}
   \end{tabular}  
\end{minipage}
\caption{The calls to the self-energy computation subroutine with the code restructure.}
\label{fig:workflow_batch}
\end{figure}
\subsection{Reducing memory allocation and data movement overhead}\label{sec:ACC1}
There are several allocatable arrays used to hold intermediate results in the self-energy and the collision integral calculations.
For example, in the MPI/OpenMPI version of the self-energy computation, the temporary array \texttt{se2tmp} that appears in Fig.~\ref{fig:OMP} is used to hold intermediate self-energy contraction results produced by each thread. This array is only used inside the subroutine which computes $\Sigma^<$ at a given $(t,t')$ pair. Therefore, it is allocated and deallocated within such a subroutine. The repeated allocation and deallocation does not introduce much overhead on CPUs. However, such overhead cannot be ignored on GPUs.

% \begin{figure}[ht!]
% \centering
% \begin{minipage}{1\linewidth}
%     \footnotesize
%     \begin{tabular}{c}
%     \begin{lstlisting}
% allocate(P(nk,dim,dim))
% allocate(se1(my_nk,dim,dim))
% allocate(se2(my_nk,dim,dim))
% P(:,:,:) = (0d0,0d0)
% se1(:,:,:) = (0d0,0d0)

% !$acc kernels loop 
% (implicitly copy(P) and copyin(GTtmp1,GTtmp2))
% ...
% !$acc end kernels

% !$acc kernels loop 
% (implicitly copy(se1) and copyin(P,GTtmp1))
% ...
% !$acc end kernels

% deallocate(P)
% se2(:,:,:)=(0d0,0d0)
   
% !$acc kernels loop 
% (implicitly copy(se2) and copyin(GTtmp1,GTtmp2))
% ...
% !$acc end kernels
% se = (factor * se1 - se2) / nk**2
% deallocate(se1)
% deallocate(se2)
%     \end{lstlisting}
%     \end{tabular}
%     \caption{The memory management in the MPI/OpenMP implementation.}
%     \label{fig:memory}
% \end{minipage}
% \end{figure}

To reduce this overhead, we modified the code to allocate these arrays only once on the host, prior to calling the self-energy calculation subroutine. %outside the loop over \texttt{itb} in Fig.~\ref{fig:workflow_batch}. 
Once allocated, these arrays are immediately copied to the GPU device. There is no need for repeated allocation and data transfer inside the loop over $t$ or $t'$. 

%use \texttt{!\$acc enter data copyin} to copy the data in the array to the GPU device only once before looping over $(t,t')$. The \texttt{!\$acc exit data delete} is used to release the memory after we exit from the loop.
The self-energy computational kernels are launched inside the subroutine \texttt{get\_2B\_SE\_kernel} as shown in Fig.~\ref{fig:ACC}. We use \texttt{!\$acc update} to update these arrays either on the host or on the device outside the kernels. The memory allocated for these arrays is released after we exit from the main loop. This memory management scheme is much faster than frequent data allocation, transfer and deallocation. 
%Each time we call the subroutine which computes the self-energy at a certain $(t,t')$ pair inside the loop, we need to initialize these arrays to $0$, and use \texttt{!\$acc update device} to update their values on device.

%\JY{commented the following codes for the main loop with OpenACC implementation}

Moreover, since we use \texttt{Gtmp1} and \texttt{Gtmp2} arrays repeatedly within the kernels launched to compute~\eqref{eq:polarizability},~\eqref{eq:kernel2} and implement  Algorithm~\ref{alg:contraction2} for self-energy computation, we use \texttt{!\$acc data} to declare an OpenACC data region within the subroutine \texttt{get\_2B\_SE\_kernel} in Fig.~\ref{fig:ACC}. The use of such an OpenACC directive eliminates the need for repeated movement of arrays between the host and device, thereby further reducing the data transfer cost.

For the test problem, optimizing memory allocation and data transfer between the host and device yields $8.78\%$ performance improvement in the self-energy computation.  Similar techniques can be used to
reduce memory allocation and data transfer overhead in the evaluation of the collision integral.

\subsection{Trading on-the-fly index calculation for reduced irregular memory access}\label{sec:ACC3}

In the OpenMP implementation shown in Fig.~\ref{fig:OMP}, the $k$-point index \texttt{iqk} over which the Green's function contractions are performed in the self-energy evaluation are retrieved indirectly from arrays  \texttt{allind\_kpk} and \texttt{allind\_kmk} which store the sums and differences of two given $k$-points. These arrays are derived from the FORTRAN structure \texttt{bz}. On CPUs, as long as these arrays, which are not very large, can fit in the L2 cache, retrieving indices stored in these arrays does not introduce much overhead. However, on GPUs, the indirect addressing facilitated by these indices breaks coalesced memory access pattern required to enable concurrency in fetching data in parallel. Furthermore, the use of derived data type makes it difficult to ensure data structures are correctly mapped and synchronized.  Because the indices stored in these arrays can all be computed by simple formulae, we can evaluate them on-the-fly.

In the model problem, we sample the $k$-points uniformly on the interval $[-\pi, \pi)$ to have
\begin{equation}
    k_j = -\pi + \frac{2(j-1)}{n_k}\pi, \quad j = 1,\, ...,\, n_k.
    \label{eq:kj}
\end{equation}
The sum of two sampled $k$-points falls within $[-2\pi, 2\pi)$. By taking into account the periodicity of the $k$-points, we can obtain the index of $k_{j_1}+k_{j_2}$ among all sampled $k$-points in \eqref{eq:kj}, for $j_1, j_2 = 1, ..., n_k$, which was originally stored in the two-dimensional array \texttt{bz\%allind\_kpk} as follows,
\begin{equation}
\begin{aligned}
    &\texttt{bz\%allind\_kpk}(j_1, j_2)\\
=& 
    \begin{cases}
        j_1 + j_2 + n_k/2 - 1, \quad & {\rm if}\; k_{j_1}+k_{j_2}<-\pi;\\
        j_1 + j_2 - n_k/2 - 1, \quad & {\rm if}\; -\pi\leq k_{j_1}+k_{j_2}<\pi;\\
        j_1 + j_2 - 3n_k/2 - 1, \quad & {\rm if}\; k_{j_1}+k_{j_2}\geq\pi.
    \end{cases}
\end{aligned}
\end{equation}

A similar expression can be used to obtain the index of $k_{j_1} - k_{j_2}$ among the sampled $k$-points for $j_1,j_2=1,2,...n_k$.
%Similarly, we can get the formula for \texttt{bz\%allind\_kmk} with $j_1,\, j_2 = 1, ..., n_k$ as
%\begin{equation}
%\begin{aligned}
%    &\texttt{bz\%allind\_kmk}(j_1, j_2)\\
%    =& 
%    \begin{cases}
%        j_1 - j_2 + 3n_k/2 + 1, \quad & {\rm if}\; k_{j_1}-k_{j_2}<-\pi;\\
%        j_1 - j_2 + n_k/2 + 1, \quad & {\rm if}\; -\pi\leq k_{j_1}-k_{j_2}<\pi;\\
%        j_1 - j_2 - n_k/2 + 1, \quad & {\rm if}\; k_{j_1}-k_{j_2}\geq\pi.
%    \end{cases}
%\end{aligned}
%\end{equation}

By evaluating these expressions on the fly, we can avoid unnecessary memory access that impedes concurrent executions of the self-energy and collision integral calculations. This optimization gives a $3.35\%$ improvement in the performance of self-energy computation.

\subsection{Performance assessment}
\begin{figure}
    \centering
    \includegraphics[width=0.5\textwidth]{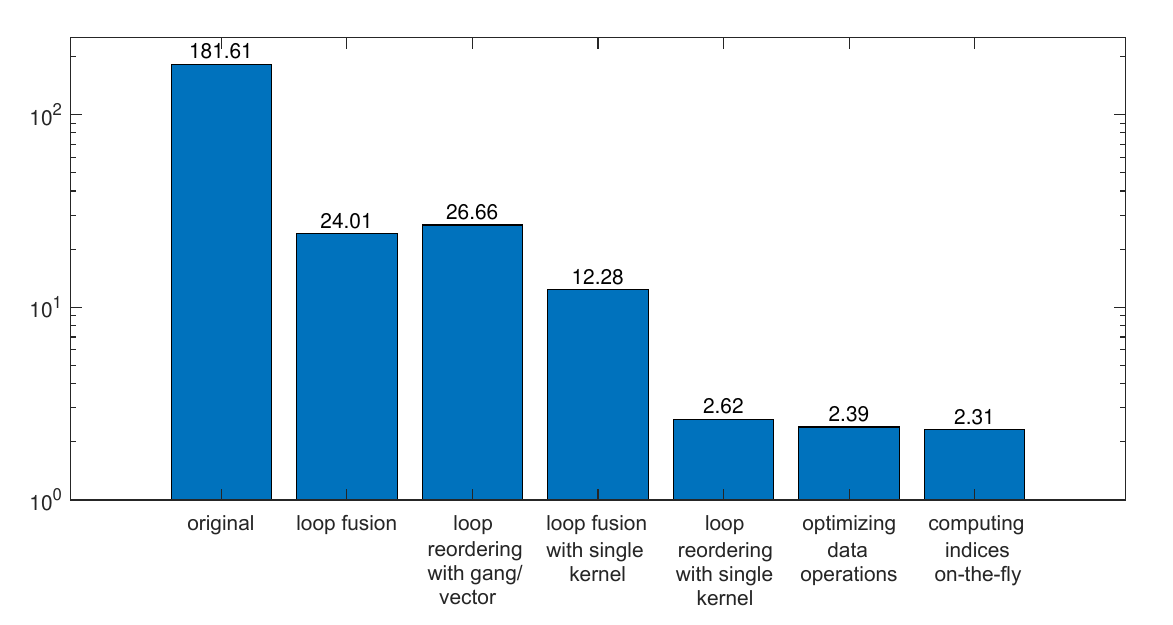}
    \caption{The wall clock time (seconds) used by different variants of the OpenACC implementation for the model problem.}
    \label{fig:acc_pinpoint}
\end{figure}

The benefits of all strategies we discussed above are displayed and compared  in Fig.~\ref{fig:acc_pinpoint}. We can see that loop reordering and the use of \texttt{loop gang} and \texttt{loop vector} (the third bar) yields the largest initial gain in performance, reducing the total wallclock time from 182 seconds to 27 seconds. The code restructure to increase the workload in kernels reduces the wall clock time by another factor of 10 (the fifth bar). Overall, we achieve a $98.73\%$ reduction in wall clock time compared to a naive OpenACC implementation after all the optimizations techniques are applied.

\begin{figure*}[htbp]
    \centering
    \includegraphics[width=0.9\textwidth]{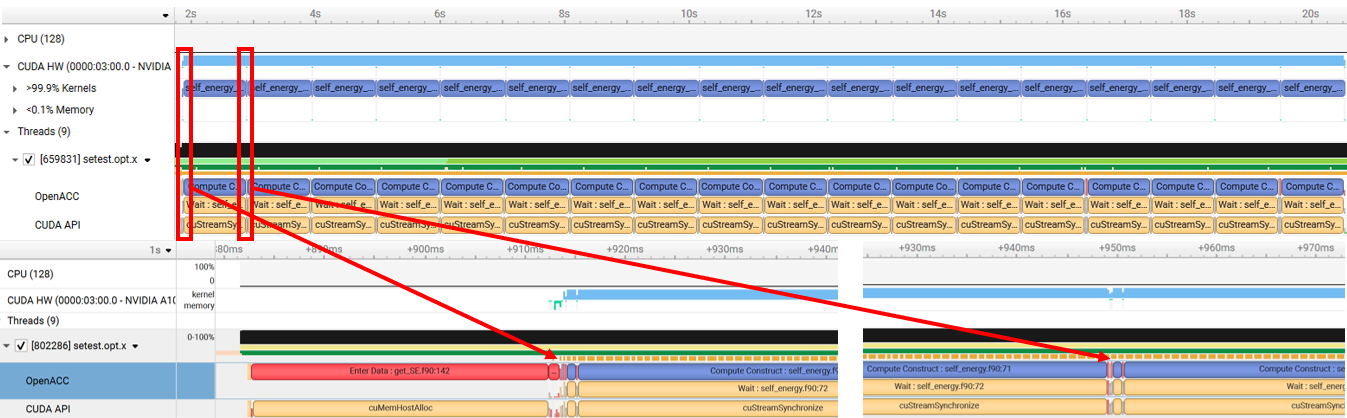}
    \caption{Performance profile generated from Nsight Systems for self-energy calculation. The top figure captures the sequence of events occurred in the entire self-energy calculation, while the bottom figure highlights work performed in the first two launched kernels.}
    \label{fig:profile_base}
\end{figure*}

\begin{figure*}[htbp]
    \centering
    \includegraphics[width=0.9\textwidth]{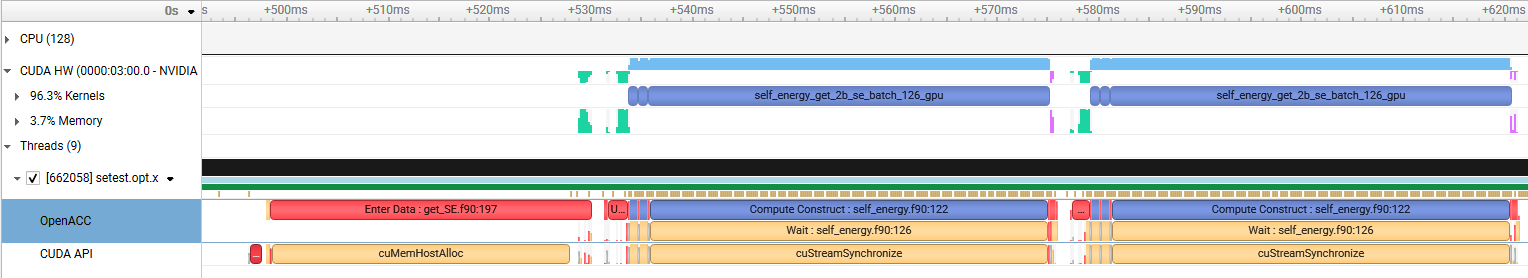}
    \caption{The profile result from Nsight Systems for the OpenACC implementations after the optimizations to compute $\Sigma(\vk;t,t')$.}
    \label{fig:profile_new}
\end{figure*}

We also used NVIDIA's Nsight Systems to generate performance profiles of the self-energy calculation before and after optimizations are applied. The Nsight Systems profiles shown in
Figures~\ref{fig:profile_base} and ~\ref{fig:profile_new} are generated from a run with $n_k=1024$ and $n_t=10$. We used $16$ MPI ranks in the computation.

Fig.~\ref{fig:profile_base} shows the profile result with direct application of the OpenACC directive \texttt{parallel loop} to the computing kernels in the self-energy calculation. We can observe that the kernels are launched frequently, and the kernel to compute $\Sigma^2$ as shown in Algorithm~\ref{alg:contraction2} costs much more than the other two kernels. It takes about $18$ seconds in total to compute the self-energy. 

After we apply all the optimizations discussed in the previous subsections, the profile result turns to Fig.~\ref{fig:profile_new}. Compared to Fig.~\ref{fig:profile_base}, there are fewer memory transfers and fewer kernel launches. It only takes about $100$ milliseconds for the calculation, which shows a significant improvement from $18$ seconds. 
%From these profile results, we have greatly improved the OpenACC implementation  from the baseline where we directly applied \texttt{!\$acc kernels}.
%

%To improve the memory access pattern, in the OpenACC implementation of the collision integral computation, we always set the time argument regarding the batching to the outermost index, and set the $k$-point index to the second outermost index. This is because the iteration over the batching index is always the outermost loop, while the iteration over the $k$-point is the second outermost one.

%With these optimizations, we manage to get hybrid MPI and OpenACC implementations which port the self-energy and collision integral calculations onto GPU and enhance their performance by utilizing the GPU computing capabilities.

\section{CUDA FORTRAN implementation}\label{sec:CUDA}
Although OpenACC provides a user friendly programming model to launch concurrent computation on GPU devices, it has some inherent limitations on the management and optimization of GPU resource utilization. Because OpenACC relies on the compiler to partition computational work loads and map them to thread blocks and threads at the compile time, programmers lack complete control over thread utilization and optimization. CUDA FORTRAN~\cite{CUDA_Fortran} provides mechanisms for programmers to explicitly define the granularity and hierarchy of concurrent execution, thereby optimizing GPU resource utilization. However, utilizing these mechanisms requires extra effort and meticulous planning. 
In this section, we show how we use CUDA FORTRAN to implement the self-energy computation. 
Similar techniques can be used for the collision integral calculations as well.

\subsection{Kernel launch}
One advantage of CUDA FORTRAN is the direct control it provides over GPUs thread utilization, allowing for customized GPU kernels optimized for performance.  A kernel launch requires one to specify the number of thread blocks and the number of threads contained in each block.  In CUDA FORTRAN, thread block and thread indices can be organized on a configurable three-dimension grid so that one can easily designate a subset of thread blocks or threads to perform a certain task while other thread blocks or threads can be used to perform other tasks concurrently. We have the flexibility to tailor the grid and block dimensions for each kernel and specify the level of parallelism for individual indices. 

Upon analyzing data dependencies, we determine that it is feasible to launch three kernels when calling the  subroutine \texttt{get\_2B\_SE\_kernel} for self-energy calculation. 
The first kernel handles the computation described in~\eqref{eq:polarizability}, the second carries out the contraction of $P$ and $G^{<}$ defined in~\eqref{eq:kernel2} as well as the computation of $\Sigma^2$ shown in Algorithm~\ref{alg:contraction2} concurrently. The last kernel sums up the first and second terms in~\eqref{eq:2BSE} by combining results produced from the preceding kernels.

All the kernels in the \texttt{get\_2B\_SE\_kernel} subroutine, are launched by specifying a number of thread blocks and threads in the \texttt{<<<blocks,threads>>>} clause as shown in Figure~\ref{fig:CUDA_control} with  both \texttt{blocks} (representing thread blocks) and \texttt{thread} 
 (representing threads within each thread block) arranged as three-dimensional grids. 
\begin{figure}[ht!]
\centering
\begin{minipage}{0.97\linewidth}
\footnotesize  
    \begin{tabular}{c}
    \begin{lstlisting}  
...
integer, parameter :: THREADS_PER_BLOCK=128
...
! kernel 2
blocks = dim3(my_nk,batch_size,2)
threads = dim3(THREADS_PER_BLOCK,1,1)
call get_2B_SE_batch_kernel<<<blocks,threads>>>&
          (se_d, ...)
...
    \end{lstlisting}
    \end{tabular}  
\end{minipage}
\caption{CUDA FORTRAN kernel launch for self-energy calculation with thread block and thread configurations specified by 3D arrays \texttt{blocks} and \texttt{threads}.}
\label{fig:CUDA_control}
\end{figure}

The dimension of \texttt{block} is defined to be $\texttt{my\_nk}\times \texttt{nt}\times 2$.
In the second kernel, the third dimension of \texttt{blocks} is used to separate thread blocks into two groups. As shown in Figure~\ref{fig:CUDA}, the group of thread blocks indexed by  \texttt{gridIdx\%z=1} are designated to compute~\eqref{eq:kernel2}, whereas the group indexed by \texttt{gridIdx\%z=2} is scheduled to compute $\Sigma^2$ at the same time. It is worth noting that OpenACC does not support this simultaneous computation of~\eqref{eq:kernel2} and $\Sigma^2$.  This distinction highlights the superior parallelization capabilities offered by CUDA FORTRAN compared to OpenACC and results in enhanced performance.
\begin{figure}[ht!]

\centering
\begin{minipage}{0.97\linewidth}
\footnotesize  
\begin{tabular}{c}
\begin{lstlisting}
kk = blockIdx%x
it = blockIdx%y
blocksize = blockDim%x
red_size = MIN(blocksize, nk)       
tid = threadIdx%x

if (blockIdx%z==1) then
  ... ! compute (9)
else
  Ptmp11(tid) = (0d0,0d0)
  ...
  ikk = start_kpts+kk-1
  do kp=tid,nk,blocksize
    do kpp=1,nk
      iqk = ...
      Ptmp11(tid) = Ptmp11(tid)+Ut(it)*...
      ...
    enddo
  enddo
  call syncthreads()
  ! reduction to thread 1
  s = 1
  do while (s<red_size)
    if (MOD(tid, 2*s)==1) then
      if (tid+s<=red_size) then
        Ptmp11(tid) = Ptmp11(tid)+Ptmp11(tid+s)
        ...
      endif
    endif
    call syncthreads();
    s = s*2
  enddo
  if (tid==1) then
    se2_d(kk,1,1,it) = Ptmp11(1)
    ...
  endif 
endif
\end{lstlisting}
\end{tabular}
\end{minipage}
\caption{The CUDA Fortran implementation of the $\Sigma^2_{jm}(\mathbf{k}; t,t')$ calculation at a particular $(t,t')$ pair.}
\label{fig:CUDA}
\end{figure}
Within the group of thread blocks designated to compute $\Sigma^2(\vk,t,t')$, thread blocks indexed by different \texttt{gridIdx\%y} values are assigned to compute $\Sigma^2(\vk,t,t')$ associated with different $t$ or $t'$ values. Thread blocks with the group indexed by the same $\texttt{gridIdx\%y}$ values and \texttt{gridIdx\%z}=2  are used to execute the $k$ loop in Algorithm~\ref{alg:contraction2} in parallel. 

All threads within each thread block are used to execute the nested $j$, $m$, $q$, $k'$ loops in parallel. The dimension of \texttt{threads} is set to $\texttt{THREADS\_PER\_BLOCK} \times 1 \times 1$, where
\texttt{blockDim\%x}=\texttt{THREADS\_PER\_BLOCK} is a pre-determined and optimized thread count. In Fig~\ref{fig:CUDA_control}, it is set to 128. However, the optimial choice of $\texttt{THREADS\_PER\_BLOCK}$ depends on the size of the problem, i.e., the number of $k$-points and the number of bands.

.

\subsection{Reduction in the shared memory for summation}
As is the case in the OpenMP and OpenACC implementations of self-energy calculation, a reduction is needed to sum up all terms in $\Sigma_{jm}^2(\mathbf{k};t,t')$ from \eqref{eq:2BSE} if the arrays used to hold the $G$ and $\Sigma$ are allocated in the shared memory which has a much higher memory bandwidth. To avoid bank conflict, we unroll the \texttt{ii}  and \texttt{jj} loop and define four temporary arrays \texttt{Ptmp11}, \texttt{Ptmp12}, \texttt{Ptmp21} and \texttt{Ptmp22} to hold $\Sigma^2_{11}$, $\Sigma^2_{12}$, $\Sigma^2_{21}$ and $\Sigma^2_{22}$ separately.

%By adopting this approach, we can concurrently compute these four arrays, resulting in concurrent computation of four elements \texttt{se2\_d} which holds the values of $\Sigma^2$ on the device within each thread block. In Fig.~\ref{fig:CUDA}, we show the operations involving \texttt{Ptmp11}, and omit the details of the other arrays for simplicity.} 

%Unrolling the loops for band indices helps reduce memory access latency and improve memory bandwidth utilization

CUDA FORTRAN offers several options to perform the reduction required to sum up several arrays. The easiest  option is to use the intrinsic function \texttt{sum}, which performs reduction automatically. For example, summing up all elements of \texttt{Ptmp11} can be done by calling \texttt{sum(Ptmp11)}. When  \texttt{THREADS\_PER\_BLOCK} is set to 128, this implementation option uses 2.23 seconds to evaluate the self-energy for the model problem defined earlier.

\begin{figure}[ht!]
    \centering
    \includegraphics[width=0.9\linewidth]{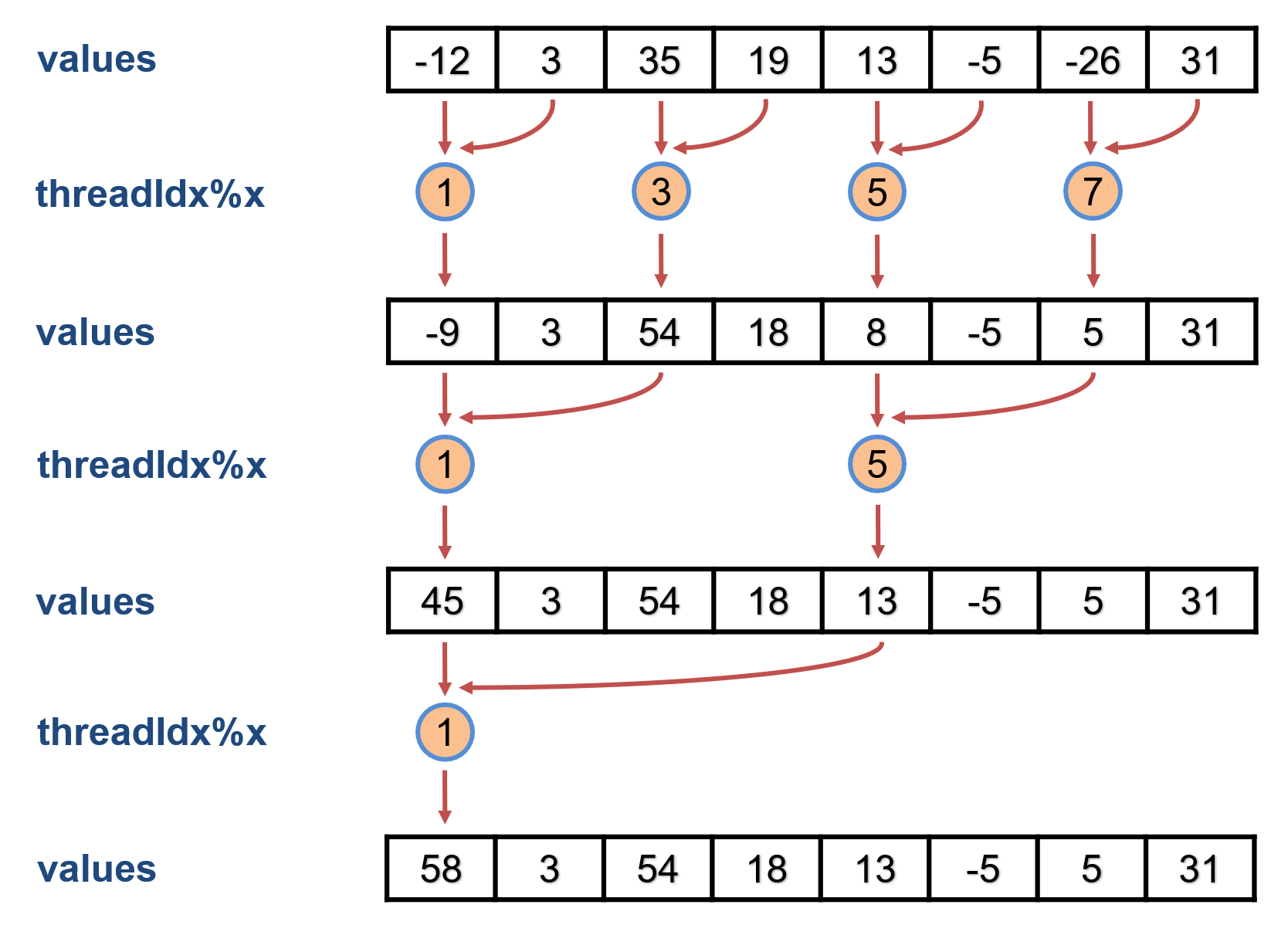}
    \caption{An illustration of a parallel binary-tree based   reduction scheme for summing up elements of an array of size $8$.}
    \label{fig:reduc}
\end{figure}

It is not clear which reduction algorithm is implemented in the \texttt{sum} function.  An efficient binary-tree based reduction algorithm can be easily implemented in CUDA FORTRAN by mapping each element of the array to be summed to a node in a binary tree, and summing these elements pairwise in a hierarchical fashion from the leaves of the tree towards the root as shown in Fig.~\ref{fig:reduc}.
One way to implement this binary-tree based reduction is to use CUDA FORTRAN's warp-level primitives to repeatedly reshuffle elements of the array to be summed to enable concurrent pairwise summation in a recursive fashion.  Fig.~\ref{fig:warpsum} shows how such a summation is performed for \texttt{Ptmp11}. At the end of the \texttt{while} loop, the first element of \texttt{Ptmp11} contains the sum of all its elements. If the \texttt{Ptmp11} has a size of \texttt{red\_size}, the number of loop iterations is
$\left\lceil \log_2(\texttt{red\_size}) \rceil\right.$ 
 -- significantly fewer than the \texttt{red\_size} iterations required for sequential summation when \texttt{red\_size} is large. Our experiments shows that takes 2.10 seconds to execute the this implementaiton for the model problem. This is slightly faster than using the \texttt{sum} function.
%The shuffle has a butterfly-type structure as shown in the following Figure~\ref{fig:shfl}
%\begin{figure}[h!]
%    \centering
%    \includegraphics[width=0.9\linewidth]{figures/shfl.png}
%    \caption{shfl\_xor example}
%    \label{fig:shfl}
%\end{figure}

\begin{figure}[ht!]
\centering
\begin{minipage}{0.97\linewidth}
\footnotesize  
    \begin{tabular}{c}
    \begin{lstlisting}  
s = 1
do while (s<red_size)
  tmp = __shfl_xor(Ptmp11, 2*(s-1)))
  Ptmp11 = Ptmp11 + tmp
  s = s*2
enddo
    \end{lstlisting}
    \end{tabular}  
\end{minipage}
\caption{A temporary array \texttt{tmp} is created by warp-level primitive \texttt{\_\_shfl\_xor} to enable recursive pairwise summation in a binary-tree based reduction algorithm.}
\label{fig:warpsum}
\end{figure}

One drawback of the implementation shown in Fig.~\ref{fig:warpsum} is that all elements of \texttt{Ptmp11} and \texttt{tmp} are summed in every loop iteration, even though the number of elements that need to be added is halved each time. In the final iteration, only \texttt{Ptmp11(1)} and \texttt{tmp(1)} need to be summed. To reduce the number of additions, we use the thread index \texttt{tid} =  \texttt{threadIdx\%x} and an offset \texttt{s}, which is set to an integer of power of 2, to select the elements of the \texttt{Ptmp11}, \texttt{Ptmp12}, \texttt{Ptmp21}, \texttt{Ptmp22} arrays to be summed pairwise in parallel within the \texttt{while} loop shown in Fig.~\ref{fig:CUDA}. A call to the \texttt{syncthread()} function is made at the end of each loop iterate to ensure all summations at each level of the binary tree are completed before we move on to the next level.  For the same test problem, this approach reduces the summation time so that the whole computation takes 1.95 seconds, which represents a 13\% improvement over the use of the intrinsic \texttt{sum} directive. {This also gives a 16\% improvement compared to the OpenACC implementation, which takes 2.31 seconds for the model problem.} 
Clearly, CUDA FORTRAN provides more flexibility to implement efficient concurrent reduction in shared memory compared to OpenACC which relies completely on the compiler for reduction. Such flexibility results in significant performance improvement.

%After the computation for \texttt{Ptmp11(tid)}, we conduct a summation reduction to \texttt{thread 1} using a \texttt{do} loop. 
%The concept of parallel reduction is illustrated in Fig.~\ref{fig:reduc} with \texttt{blockDim\%x}=\texttt{8}. 
%For a reduction with \texttt{red\_size} arguments as in Fig.~\ref{fig:CUDA}, the number of iterations is determined by $\texttt{ceil}(\log_2(\texttt{red\_size}))$. During each iteration, the summations at different \texttt{tid} values are executed concurrently. This approach significantly reduces the computational complexity associated with performing reductions. {\color{cyan}For our model problem, this implementation costs 1.07 seconds, which outperforms the other two methods previously discussed.}

\subsection{Memory management}
In CUDA FORTRAN, explicit memory management helps reduce memory allocation and data transfer overheads. Arrays can be directly allocated on the GPU if they are used exclusively on the GPU. Such an array can be passed as an argument to a subroutine and be used on the GPU within the subroutine. This feature eliminates the need to transfer data from a CPU to a GPU and update the data in a coherent fashion, which is required in OpenACC.

By allocating the arrays \texttt{se1\_d} and \texttt{se2\_d} that hold $\Sigma_{jm}^1$ and $\Sigma_{jm}^2$ respectively on the GPU directly as shown in Fig.~\ref{fig:memory_fortran} and passing them to the subroutine \texttt{get\_2B\_SE\_kernel}, we can avoid the copying and update required in the OpenACC implementation shown in Fig.~\ref{fig:ACC} and reduce memory management overhead.  
\begin{figure}[ht!]
\centering
\begin{minipage}{0.93\linewidth}
\footnotesize
\begin{tabular}{c}
\begin{lstlisting}
complex(DPC), device, allocatable ::&
    Gtmp1_d(:,:,:,:), Gtmp2_d(:,:,:,:),&
    U_t1_d(:), U_t2_d(:), P_d(:,:,:,:),&
    se1_d(:,:,:,:), se2_d(:,:,:,:)
...
allocate(Gtmp1_d(nk,dim,dim,nt))
...
allocate(se2_d(my_nk,dim,dim,nt))
...
Gtmp1_d = G(:,:,:,&
    ((itp-1)*nt+1):(itp*nt),nt)
...
call get_2B_SE_kernel(se_d, my_nk, Gtmp1_d,&
    Gtmp2_d, U_t2_d, U_t1_d, P_d, se1_d, se2_d)
se(:,:,:,((itp-1)*nt+1):(itp*nt),nt)= se_d
...
\end{lstlisting}
\end{tabular}
\end{minipage}
\caption{Device memory allocation and data transfer between host and device in CUDA FORTRAN.}
\label{fig:memory_fortran} 
\end{figure}

CUDA FORTRAN also allows data allocated on the host to be copied to the device and vice versa through a simple assignment statement. For example, in Fig.~\ref{fig:memory_fortran}, the assignment 

{\small \texttt{Gtmp1\_d = G(:,:,:,((itp-1)*nt+1):(itp*nt),nt)}}
\noindent is used to copy a portion of the array \texttt{G}  allocated on the host to the \texttt{Gtmp1\_d} array allocated on the device.  Similarly, the assignment of \texttt{se\_d} to \texttt{se(:,:,:,((itp-1)*nt+1):(itp*nt),nt)} at the end of the do loop in Fig.~\ref{fig:memory_fortran} copies data from the device to the host.

%This approach provides full control over the memory management of both the host and device, enabling us to explicitly minimize both memory allocation and data transfer. But we also notice that the complexity of the code increases when using CUDA FORTRAN. \CY{what does that mean? what complexity are we talking about?} \JY{This means in OpenACC we only need to call some directives such as \texttt{copyin} and \texttt{update device}, but in CUDA FORTRAN we have to write out all data management explicitly, so the coding becomes more complicated.}

\section{Computational results}\label{sec:results}

\subsection{The hardware}

We conduct all computational experiments on the Perlmutter supercomputer maintained at National Energy Scientific Computing Center (NERSC).
%, which has $1536$ \JY{add two together? 1536+256} GPU nodes and 3072 CPU nodes. 
%Each CPU node is composed of $2$ AMD EPYC 7763 (Milan) CPUs with $64$ cores per CPU, and $2$ threads per core. In total, there is 512 GB DDR4 memory for each node, and for each CPU, the memory bandwith is 204.8 GB/s. The compute performance is 39.2 GFlops per core.
The CPU nodes on Perlmutter were described in Section~\ref{sec:MPI}, and each GPU node on Perlmutter is composed of $1$ Milan CPU and $4$ NVIDIA A100 (Ampere) GPUs. Each GPU is equipped with 40 GB of high bandwidth memory (HBM)  with 1555.2 GB/s GPU memory bandwidth. The peak performance is 9.7 GPU TFlops for the FP64 data type. 

\subsection{Comparison of the OpenMP, OpenACC and CUDA FORTRAN implementations}

In this subsection, we compare the performance of the self-energy as well as the collision integral computation implemented with OpenMP (for CPUs), OpenACC and CUDA FORTRAN (for GPUs). CUDA aware MPI is used to for internode communication. 

In the first example, we set $n_k=1024$, $n_t=500$ to ensure that a significant amount of work is performed in these calculations and the performance of these calculations can be measured accurately. We then vary the number of MPI ranks used to parallelize the computation among different nodes.

The OpenMP implementation is always run with $64$ threads within each MPI rank, $4$ MPI ranks are launched on each CPU node. 
%For the GPU implementations, we set $\texttt{batch\_size}=256$ which is sufficiently large to observe the impact of batching.

The performance of different implementations for self-energy and collision integral calculations is presented in Table~\ref{tab:test1} and Table~\ref{tab:test1_CI}, respectively.

\begin{table*}[!ht]
\renewcommand{\arraystretch}{1.3}
\caption{Comparison of wall clock time (in seconds) taken to perform self-energy calculations in parallel on CPUs and GPUs using different programming models. These calculations use $n_k=1024$ k-points and $n_t=500$ time steps.}
\label{tab:test1}
    \centering
    \begin{tabular}{|c|p{1.5cm}<{\centering}|p{1.5cm}<{\centering}|p{1.5cm}<{\centering}|p{1.5cm}<{\centering}|p{1.5cm}<{\centering}|p{1.5cm}<{\centering}|p{1.5cm}<{\centering}|}
       \hline
       \multirow{2}*{} & \multicolumn{7}{c|}{number of MPI ranks} \\
       \cline{2-8}
       &  $16$  & $32$  & $64$  & $128$  & $256$  & $512$  & $1024$ \\
       \hline
       %MPI (CPU) & $31895.55$ & $16202.51$ & $8146.84$ & $4260.47$ & $2283.65$ & $41312.80$ & $813.80$\\
       %\hline
       OpenMP (CPU) & $1589.12$ & $816.04$ & $438.65$ & $204.39$ & $164.28$ & $117.00$ & $97.62$\\
       \hline
       %\textbf{Speedup} & $20.07$ & $19.86$ & $18.57$ & $20.84$ & $13.90$ & $11.22$ & $8.34$\\
       %\hline
       %OpenACC (GPU) & $11.46$ & $6.10$ & $3.39$ & $2.05$ & $1.40$ & $1.05$ & $1.06$ \\ % use batching with batch size=256
       % \hline
       % \textbf{Speedup} & $138.67$ & $133.78$ & $129.40$ & $99.70$ & $117.34$ & $111.43$ & $92.09$\\
       % \hline
       OpenACC (GPU) & $11.21$ & $5.96$ & $3.32$ & $1.99$ & $1.33$ & $1.03$ & $0.91$ \\
       \hline
       \textbf{Speedup} & $141.76$ & $136.92$ & $132.12$ & $102.71$ & $123.52$ & $113.59$ & $107.27$\\
       \hline
       % CUDA FORTRAN (GPU) & $8.45$ & $4.53$ & $2.58$ & $1.57$ & $1.12$ & $0.95$ & $0.84$\\ % use batching with batch size=256
       % \hline      
       % \textbf{Speedup} & $188.06$ & $180.14$ & $170.02$ & $130.18$ & $146.68$ & $123.16$ & $116.21$\\
       % \hline
       CUDA FORTRAN (GPU) & $9.06$ &  $4.84$ & $2.72$ & $1.66$ & $1.13$ & $0.94$ & $0.77$\\
       \hline      
       \textbf{Speedup} & $175.40$ & $168.60$ & $161.27$ & $123.13$ & $145.38$ & $124.47$ & $126.78$\\
       \hline
    \end{tabular}
\end{table*}

\begin{table*}[!ht]
\renewcommand{\arraystretch}{1.3}
\caption{Comparison of wall clock time (in seconds) taken to perform the collision integral calculation in parallel on CPUs and GPUs using different programming models. These calculations use $n_k=1024$ k-points and $n_t=500$ time steps.}
\label{tab:test1_CI}
    \centering
    \begin{tabular}{|c|p{1.5cm}<{\centering}|p{1.5cm}<{\centering}|p{1.5cm}<{\centering}|p{1.5cm}<{\centering}|p{1.5cm}<{\centering}|p{1.5cm}<{\centering}|p{1.5cm}<{\centering}|}
       \hline
       \multirow{2}*{} & \multicolumn{7}{c|}{number of MPI ranks} \\
       \cline{2-8}
       &  $16$  & $32$  & $64$  & $128$  & $256$  & $512$  & $1024$ \\
       \hline
       % MPI (CPU) & $197.09$ & $88.88$ & $45.92$ & $24.58$ & $11.46$ & $5.44$ & $2.80$\\
       % \hline
       OpenMP (CPU) & $124.93$ & $74.89$ & $31.25$ & $19.93$ & $9.42$ & $5.06$ & $2.84$\\
        \hline
       % OpenACC (GPU) & $10.05$ & $6.30$ & $2.63$ & $1.52$ & $0.70$ & $0.38$ &  $0.22$\\ % use batching with batch size = 256
       % \hline
       % \textbf{Speedup} & $12.43$ & $11.89$ & $11.88$ & $13.11$ & $13.46$ & $13.32$ & $12.91$\\
       % \hline
       OpenACC (GPU) & $7.83$ & $3.95$ & $2.03$ & $1.04$ & $0.56$ & $0.27$ & $0.16$ \\
       \hline
       \textbf{Speedup} & $15.96$ & $18.96$ & $15.39$ & $19.16$ & $16.82$ & $18.74$ & $17.75$\\
       \hline
    \end{tabular}
\end{table*}

%From the first row in Tabel~\ref{tab:test1}, we observe that with a doubled MPI rank, the computational time is reduced by half. This shows good strong scaling of the MPI implementation, which serves as a good baseline for comparison.

In Table~\ref{tab:test1}, we report both the wall clock times used by each implementation to compute the self-energy and the speedup achieved by the OpenACC and CUDA FORTRAN implementations executed on GPUs relative to the OpenMP implementation executed on CPUs. We observe that both the OpenACC and CUDA FORTRAN implementations executed on GPUs are more than $100$x faster than the OpenMP implementation executed on CPUs. The CUDA FORTRAN implementation consistently outperforms the OpenACC implementation, which underscores the advantage of having full control over the memory management and kernel launch. On both CPUs and GPUs, the self-energy computation has nearly perfect strong MPI parallel scaling up to 128 MPI ranks. Performance improvement can still be observed with 1024 MPI ranks. However, the scaling beyond 128 MPI ranks is less than perfect. This is due to increased MPI communication overhead relative to reduced computational workload on each MPI rank. Because the computational time consumed on each GPU is significantly lower than that on a CPU, communication overhead becomes more pronounced in GPU runs. As a result, the OpenACC and CUDA FORTRAN speedup over the OpenMP implementation also decreases slightly in general.

The performance of the collision integral calculation is reported in Table~\ref{tab:test1_CI}. We observe that the OpenACC implementation of the collision integral calculation executed on GPUs is consistently 15--19 times faster than the OpenMP implementation executed on CPUs. Because the amount of work in the collision integral calculation is much less than for that in the self-energy calculation, we cannot take full advantage of the concurrency provided by a large number of thread blocks and threads. When $n_t=500$, the collision integral calculation time is significantly less than that required for self energy calculation on CPUs. However, when ported to GPUs, the collision integral calculation time is close to that used for self-energy calculation. Furthermore, excellent strong scaling is observed for collision integral calculation with up to 512 MPI ranks in the OpenACC implementation. Communication overhead contributed to a slight slow down when the computation is performed on 1024 MPI ranks.

In addition to strong scaling, we also examined the weak scaling of the GPU implementations of the self-energy and collision integral calculation by fixing the number of $k$-points on each MPI rank to 16. As a result, in each run the total number of $k$-points which defines the size of the problem to be solved is set to $n_k=16\times$ the number of MPI ranks. For this experiment, we keep the values of $n_t$ unchanged.
%The performance results for self-energy and collision integral can be found in Tables~\ref{tab:test2} and~\ref{tab:test2_CI}, respectively.

As we discussed in section~\ref{sec:kernels}, the overall complexity of the self-energy calculation is $\mathcal{O}(n_k^3)$. However, the evaluation of $\Sigma^1$ only takes $\mathcal{O}(n_k^2)$ operations. As a result, in the weak scaling experiments, when the number of MPI ranks is doubled, we expect the wallclock time to increase by a factor that is between 2 and 4.  Such scaling is indeed observed in Table~\ref{tab:test2}.  We also observe that,  when the number of MPI ranks is fewer than 8, the OpenACC implementation takes significantly more time than the CUDA FORTRAN version. This performance gap is likely due to suboptimal thread block and thread configurations selected by the compiler during the launch of kernels in OpenACC-decorated loops.
%Better scaling is observed in the OpenMP implementation for CPUs, as there is another parallelism for the $k$-points inside the subroutine \CY{not sure what this means?}. IGPU implementations, when the number of MPI ranks is larger than $8$, the weak scaling ratio is almost $4$. This is probably because the total thread number on each MPI rank does not change. \CY{not sure what this means} Notably, we observe that MPI rank $=4$ to MPI rank $=8$, there is a reduction in the computational time for the OpenACC implementation. \CY{not sure what we should expect.} This can be attributed to the reliance of OpenACC on the compiler to determine the parallelism structure. Without manually setting the gang and vector sizes, the compiler assigns a vector size of $128$ for this specific problem. Consequently, the OpenACC version performs optimally when there are $128$ $k$-points in total taking into account the thread management overhead. \CY{not clear what this means. 128 $k$-points spread over 8 MPI ranks. the number of $k$-points per GPU doesn't change.} \JY{But when we compute the self-energy, there are still two loops over all $k$-points as shown in Fig. 3.} However, even in this case, the CUDA FORTRAN implementation still outperforms the OpenACC implementation.
\begin{table*}[!t]
\renewcommand{\arraystretch}{1.3}
\caption{The wall clock time (in seconds) of the self-energy calculation for problems in which  $n_k$ is set to 16 times the number of MPI ranks on CPUs and GPUs. The number of time steps is set to $n_t=500$.}
\label{tab:test2}
    \centering
    \begin{tabular}{|c|p{1.5cm}<{\centering}|p{1.5cm}<{\centering}|p{1.5cm}<{\centering}|p{1.5cm}<{\centering}|p{1.5cm}<{\centering}|p{1.5cm}<{\centering}|p{1.5cm}<{\centering}|}
       \hline
       \multirow{2}*{} & \multicolumn{7}{c|}{number of MPI ranks} \\
       \cline{2-8}
       &  $1$  & $2$  & $4$  & $8$  & $16$  & $32$  & $64$ \\
       \hline
       % MPI (CPU) & $2.08$ & $8.04$ & $31.54$ & $128.13$ & $515.74$ & $2040.89$ & $8146.84$ \\
       % \hline
       % OpenMP (CPU) & $69.37$ & $76.16$ & $76.35$ & $89.21$ & $117.16$ & $185.59$ & $438.65$ \\
       % \hline
       OpenMP (CPU) & $1.91$ & $3.68$ & $7.97$ & $19.87$ & $52.96$ & $185.59$ & $438.65$ \\
       \hline
       % \textbf{Speedup} & $0.03$ & $0.11$ & $0.41$ & $1.44$ & $4.40$ & $11.00$ & $18.57$ \\
       % \hline
       % OpenACC (GPU) & $0.42$ & $0.41$ & $1.04$ & $0.11$ & $0.28$ & $1.15$ & $3.39$ \\ % batched version
       % \hline
       % \textbf{Speedup} & $4.55$ & $8.98$ & $7.66$ & $180.64$ & $189.14$ & $161.38$ & $129.40$ \\
       % \hline
       OpenACC (GPU) & $0.49$ & $0.43$ & $1.00$ & $0.08$ & $0.25$ & $0.91$ & $3.33$ \\
       \hline
 %      \textbf{Speedup} &  &  &  &  &  &  &  \\
 %      \hline
       % CUDA Fortran (GPU) & $0.02$ & $0.02$ & $0.04$ & $0.09$ & $0.24$ & $0.87$ & $2.58$\\ % batched version
       % \hline      
       % \textbf{Speedup} & $95.50$ & $184.00$ & $199.25$ & $220.78$ & $220.67$ & $213.32$ & $170.02$ \\
       % \hline
       CUDA Fortran (GPU) & $0.01$ & $0.02$ & $0.03$ & $0.09$ & $0.24$ &  $0.78$ & $2.73$\\
       \hline      
 %      \textbf{Speedup} &  &  &  &  &  &  &  \\
 %      \hline
    \end{tabular}
\end{table*}

\begin{table*}[!t]
\renewcommand{\arraystretch}{1.3}
\caption{The wall clock time (in seconds) of the collision integral calculation for problems in which  $n_k$ is set to 16 times the number of MPI ranks on CPUs and GPUs. The number of time steps is set to $n_t=500$.}
\label{tab:test2_CI}
    \centering
    \begin{tabular}{|c|p{1.5cm}<{\centering}|p{1.5cm}<{\centering}|p{1.5cm}<{\centering}|p{1.5cm}<{\centering}|p{1.5cm}<{\centering}|p{1.5cm}<{\centering}|p{1.5cm}<{\centering}|}
       \hline
       \multirow{2}*{} & \multicolumn{7}{c|}{number of MPI ranks} \\
       \cline{2-8}
       &  $1$  & $2$  & $4$  & $8$  & $16$  & $32$  & $64$ \\
       \hline
       % MPI (CPU) & $43.67$ & $44.12$ & $44.16$ & $44.00$ & $45.34$ & $47.19$ & $45.92$ \\
       % \hline
       OpenMP (CPU) & $28.72$ & $29.09$ & $30.41$ & $31.50$ & $30.62$ & $31.56$ & $31.25$ \\
       \hline
       %OpenACC (GPU) & $3.84$ & $3.65$ & $3.81$ & $3.31$ & $3.46$ & $2.56$ & $2.63$ \\
       %\hline
       OpenACC (GPU) & $2.55$ & $2.91$ & $3.52$ & $2.17$ & $2.20$ & $2.21$ & $2.22$\\
       \hline
       % OpenACC (test1) & $2.50$ & $2.85$ & $3.53$ & $2.17$ & $3.81$ & $2.21$ & $2.23$ \\
       % \hline
       % OpenACC (test2) & $2.55$ & $2.88$ & $3.49$ & $2.17$ & $2.21$ & $2.22$ & $2.24$ \\
       % \hline
       % OpenACC (test3) & $2.60$ & $3.01$ & $3.55$ & $2.18$ & $2.19$ & $2.20$ & $2.19$\\
       % \hline
       %\textbf{Speedup} & $11.26$ & $10.00$ & $8.64$ & $14.52$ & $13.92$ & $14.28$ & $14.08$\\
       \hline
    \end{tabular}
\end{table*}

Because the complexity of collision integral evaluation is $\mathcal{O}(n_k)$, we expect the wall clock time to remain close to constant when we double the number of MPI ranks. Such scaling is indeed observed in  Table~\ref{tab:test2_CI}.
%The computational time for both implementations remains relatively stable across various MPI ranks. This stems from the fact that the workload assigned to each MPI rank is almost the same with a fixed \texttt{my\_nk}. \CY{what is the expected weak scaling ratio?} \JY{I think it is $1$.}

\subsection{Tuning the grid and block sizes in OpenACC and CUDA FORTRAN}

\begin{figure}[!t]
    \centering
    \includegraphics[width=0.9\linewidth]{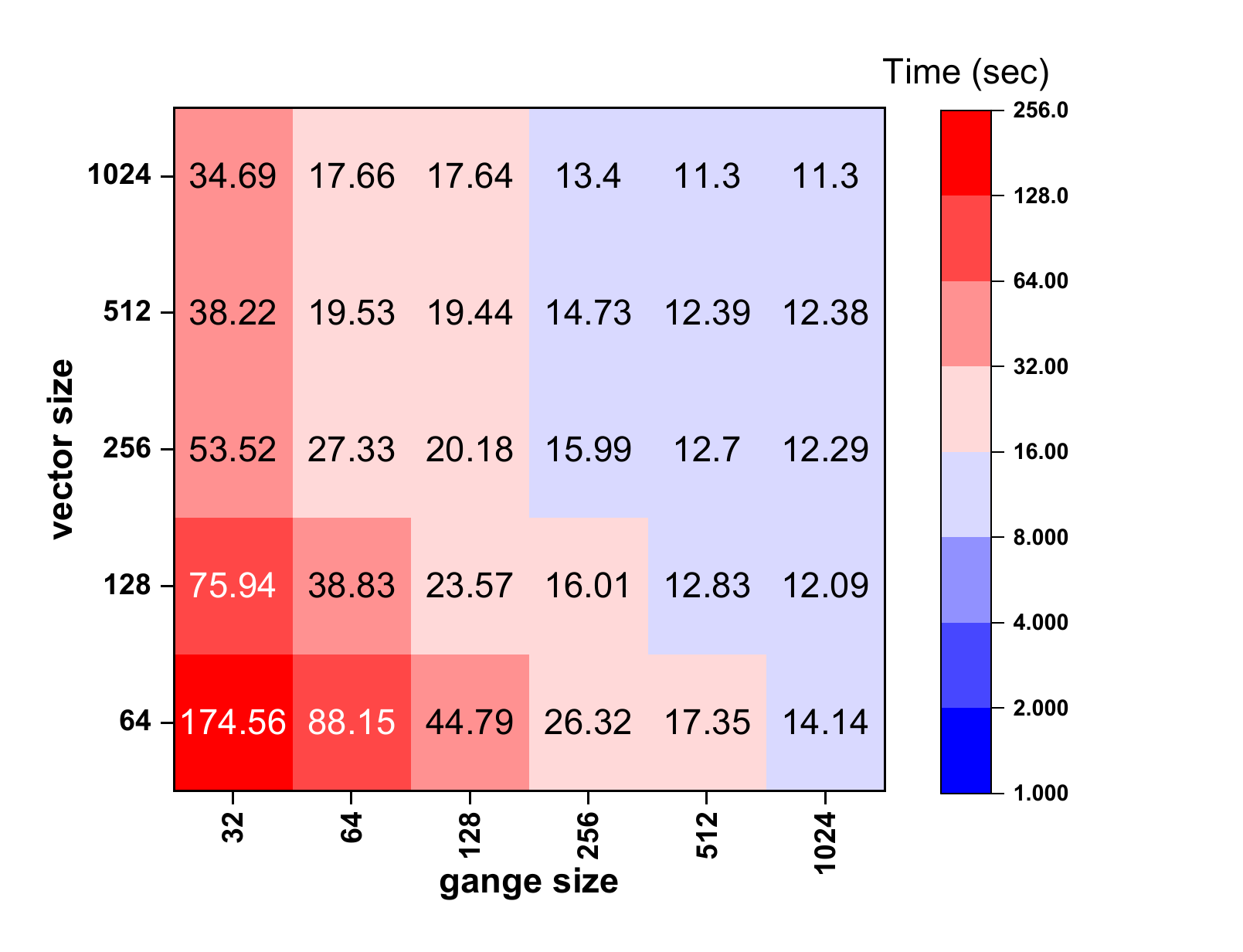}
    \caption{Comparison of wall clock time (in seconds) used in the OpenACC implementation of the self-energy calculation with different gang and vector sizes. These calculations use $n_k = 1024$ k-points and $n_t=500$ time steps, and are run on $16$ MPI ranks.}
    \label{fig:test3}
\end{figure}

\begin{figure}[!t]
    \centering
    \includegraphics[width=0.9\linewidth]{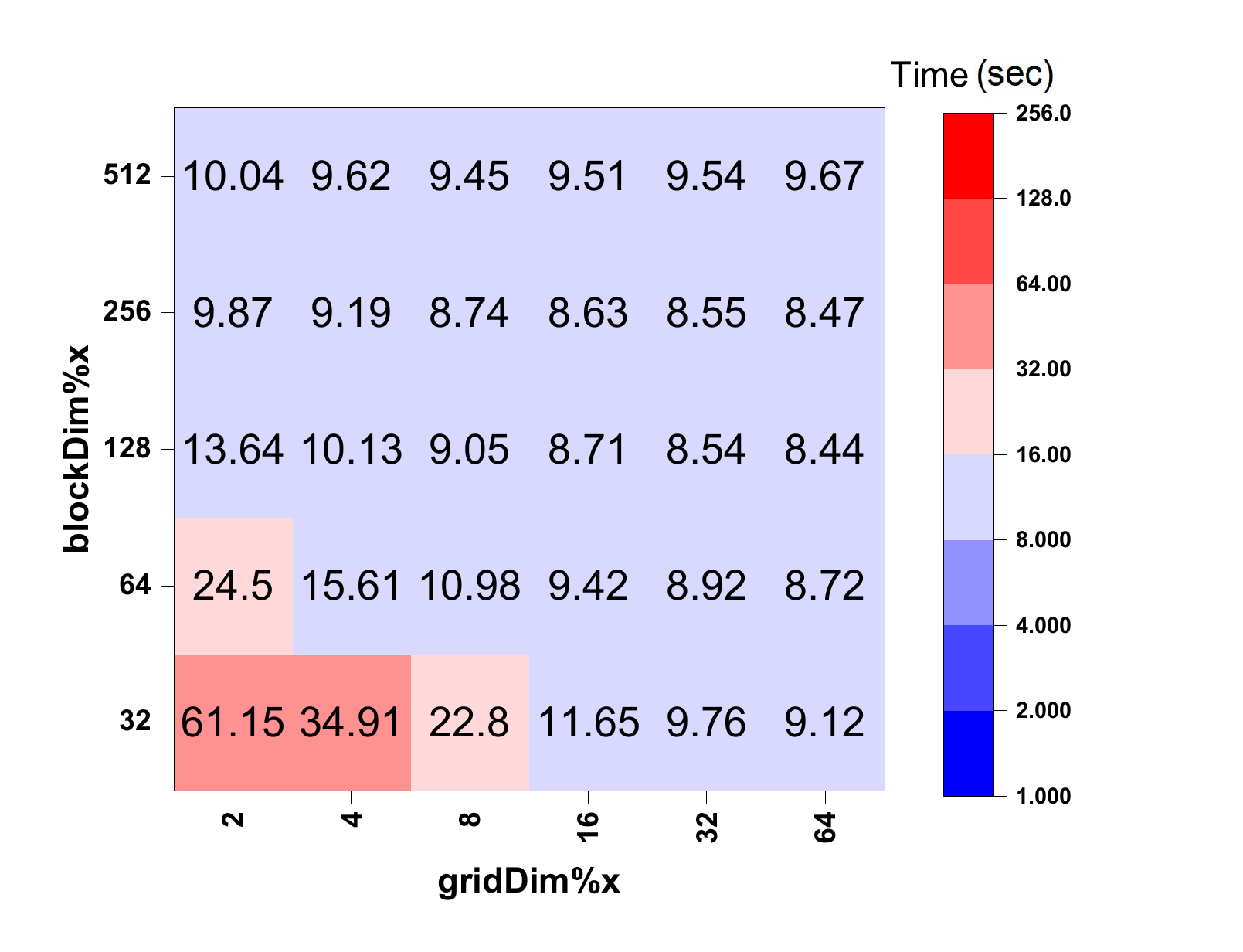}
    \caption{
    Comparison of wall clock time (in seconds) used in the CUDA FORTRAN implementation of the self-energy calculation with different thread grid and block sizes. These calculations use $n_k = 1024$ k-points and $n_t=500$ time steps, and are run on $16$ MPI ranks.
    }
    \label{fig:test4}
\end{figure}

The performance of OpenACC and CUDA FORTRAN implementations varies with respect to several parameters such as the gang and vector size in OpenACC and dimension of the thread block and thread grids. In this subsection, we examine how different choices of these parameter affect the performance of self-energy calculation. The test results presented below uses $n_k=1024$, $n_t=500$. The computation is distributed among 16 MPI ranks. The wall clock time associated with different choices of vector size and gangle size in OpenACC in shown in  Fig.~\ref{fig:test3} as a heatmap. Fig.~\ref{fig:test4} shows the wall clock time of self-energy calculation used by the CUDA FORTRAN implementation for different choices of blockDim and gridDim.  These figures illustrate the impact of various gang sizes and vector sizes in OpenACC, as well as different grid and block dimensions in CUDA FORTRAN, respectively.
% \begin{table*}[!t]
% \renewcommand{\arraystretch}{1.3}
% \caption{Comparison of running times with $n_k = 1024$, $n_t=500$, $\texttt{batch\_size}=256$, $16$ MPI ranks and different gang and vector sizes in the OpenACC implementation for computing the self-energy. The results are represented in seconds.}
% \label{tab:test3}
%     \centering
%     \begin{tabular}{|c|p{1.5cm}<{\centering}|p{1.5cm}<{\centering}|p{1.5cm}<{\centering}|p{1.5cm}<{\centering}|p{1.5cm}<{\centering}|p{1.5cm}<{\centering}|p{1.5cm}<{\centering}|}
%        \hline
%        \multicolumn{2}{|c|}{\multirow{2}*{}}  & \multicolumn{6}{c|}{gang size} \\
%        \cline{3-8}
%        \multicolumn{2}{|c|}{}   & $32$  & $64$  & $128$  & $256$  & $512$  & $1024$ \\
%        \hline
%        \multirow{5}*{vector size}  & $64$ & $174.56$ & $88.15$ & $44.79$ & $26.32$ & $17.35$ & $14.14$\\
%        \cline{2-8}
%         & $128$ & $75.94$ & $38.83$ & $23.57$ & $16.01$ & $12.83$ & $12.09$ \\
%        \cline{2-8}
%         & $256$ & $53.52$ & $27.33$ & $20.18$ & $15.99$ & $12.70$ & $12.29$ \\
%        \cline{2-8}
%         & $512$ & $38.22$ & $19.53$ & $19.44$ & $14.73$ & $12.39$ & $12.38$\\
%         \cline{2-8}
%         & $1024$ & $34.69$ & $17.66$& $17.64$ & $13.40$ & $11.30$ & $11.30$ \\
%        \hline    
%     \end{tabular}
%     %\includegraphics[]{figures/ACC_TUNING.pdf}
% \end{table*}

% \begin{table*}[!t]
% \renewcommand{\arraystretch}{1.3}
% \caption{Comparison of running times with $n_k = 1024$, $n_t=500$, $\texttt{batch\_size}=256$, $16$ MPI ranks and different grid and block sizes in the CUDA FORTRAN implementation for computing the self-energy. The results are represented in seconds.}
% \label{tab:test4}
%     \centering
%     \begin{tabular}{|c|p{1.5cm}<{\centering}|p{1.5cm}<{\centering}|p{1.5cm}<{\centering}|p{1.5cm}<{\centering}|p{1.5cm}<{\centering}|p{1.5cm}<{\centering}|p{1.5cm}<{\centering}|}
%        \hline
%        \multicolumn{2}{|c|}{\multirow{2}*{}}  & \multicolumn{6}{c|}{\texttt{gridDim\%x}} \\
%        \cline{3-8}
%        \multicolumn{2}{|c|}{}   & $2$  & $4$  & $8$  & $16$  & $32$  & $64$ \\
%        \hline
%        \multirow{5}*{\texttt{blockDim\%x}}& $32$ & $61.15$ & $34.91$ & $22.80$ & $11.65$ & $9.76$ & $9.12$\\
%        \cline{2-8}
%         & $64$ & $24.50$ & $15.61$ & $10.98$ & $9.42$ & $8.92$ & $8.72$\\
%        \cline{2-8}
%         & $128$ & $13.64$ & $10.13$ & $9.05$ & $8.71$ & $8.54$ & $8.44$\\
%        \cline{2-8}
%         & $256$ & $9.87$ & $9.19$ & $8.74$ & $8.63$ & $8.55$ & $8.47$ \\
%        \cline{2-8}
%         & $512$ & $10.04$ & $9.62$ & $9.45$ & $9.51$ & $9.54$ & $9.67$\\    
%        \hline    
%     \end{tabular}
% \end{table*}

From these figures, a general trend can be observed where the computational time tends to decrease as gang and vector sizes in OpenACC, as well as grid and block dimensions in CUDA FORTRAN, increase. 

In the OpenACC implementation, the choice of the gang size exhibits a more significant impact on the result. After increasing the gang size to $1024$, there is little difference when we increase the vector size from $128$ to $512$. Surprisingly, a further improvement in performance is achieved when using a vector size of $1024$. Based on the data presented in Fig.~\ref{fig:test3}, optimal choices for this example appear to be a gang size of $512$, and a vector size of $1024$.

In the CUDA FORTRAN implementation, the key grid and block dimensions to consider are \texttt{gridDim\%x} and \texttt{blockDim\%x}. Given the current problem setting, \texttt{gridDim\%x} is limited to a maximum of $64$ since $\texttt{my\_nk}=64$. Analysis of the results in Fig.~\ref{fig:test4} reveals that once we increase \texttt{blockDim\%x} to $128$, further increasing it do not lead to noticable improvements. The performance even becomes worse when increasing it to $512$, possibly due to excessive GPU resource requirements or significant thread divergence. The limited shared memory restricts setting \texttt{blockDim\%x} to $1024$ as with the vector size in OpenACC.
The optimal choice, based on the results in Fig.~\ref{fig:test4}, is to set $\texttt{gridDim\%x}=64$ and $\texttt{blockDim\%x}=128$. This choice results in a $1.34$x speedup when compared to the best computational time achieved with the OpenACC implementation.

\section{Conclusion}\label{sec:conclusion}
In this work, we developed high-performance implementations of a numerical method for solving the Kadanoff-Baym equations (KBE) on both CPU and GPU architectures, using a range of parallel programming models. For GPU-based acceleration, we developed several optimization techniques aimed at improving performance by maximizing computational resource utilization and minimizing overhead associated with kernel launches and memory management.

Our experiments demonstrate that the GPU implementation of the self-energy computation, developed using both OpenACC and CUDA FORTRAN, achieves over 100× speedup on NVIDIA A100 GPUs compared to an AMD EPYC many-core CPU baseline. For the collision integral computation, GPU acceleration yields more than a 10× performance gain. We found that the explicit control over thread hierarchy and memory enabled by CUDA FORTRAN leads to better performance than OpenACC, highlighting the value of fine-grained optimization in GPU programming.

All three implementations (MPI/OpenMP CPU, MPI/OpenACC GPU, and MPI/CUDA FORTRAN GPU) exhibit excellent strong and weak parallel scaling, underscoring the portability and scalability of our approach across modern high-performance computing systems. These results demonstrate the viability of our method for large-scale non-equilibrium Green's function calculation for quantum many-body systems and contribute to advancing the state of the art in GPU-accelerated scientific computing.

\section*{Acknowledgment}
This work is supported by the Center for Computational Study of Excited-State Phenomena in Energy Materials (C2SEPEM) at the Lawrence Berkeley National Laboratory, which is funded by the U.\,S. Department of Energy, Office of Science, Basic Energy Sciences, Materials Sciences and Engineering Division, under Contract No. DE-AC02-05CH11231, as part of the Computational Materials Sciences Program.  This work is also supported by the U.S. Department of Energy, Office of Science, Office of Advanced Scientific Computing Research, Scientific Discovery through Advanced Computing (SciDAC) program. The authors acknowledge the computational resources of the National Energy Research Scientific Computing (NERSC) center using ASCR-ERCAP-m1027.



\bibliographystyle{IEEEtran}
\bibliography{ref}
%\begin{thebibliography}{1}


%\end{thebibliography}

\end{document}